\documentclass[a4paper,11pt]{article}
\usepackage{graphicx}
\graphicspath{{Figures/}}
\usepackage[fleqn]{amsmath}
\usepackage{amssymb}
\usepackage{amsmath}
\usepackage[mathscr]{euscript}
\usepackage{times}
\usepackage{natbib}
\usepackage[margin=3cm]{geometry}

\title{Geophysical inversion and Optimal Transport}
\date{}
\usepackage[noblocks]{authblk}

\author[1]{Malcolm Sambridge\footnote{E-mail: malcolm.sambridge@anu.edu.au}}
\author[2]{Andrew Jackson}
\author[1,3]{Andrew P.~Valentine}
\affil[1]{Research School of Earth Sciences, The Australian National University, Canberra ACT 2601, Australia.}
\affil[2]{Institute for Geophysics, ETH Zurich, Sonneggstrasse 5, 8092 Z\"{u}rich, Switzerland.}
\affil[3]{Dept. of Earth Sciences, Durham University, South Road, Durham, DH1 3LE, UK.}
\usepackage{hyperref}
\begin{document}
\maketitle
\vspace{-40pt}

\begin{center}
\fbox{\parbox{13cm}{\small This is a pre-copyedited, author-produced PDF of an article accepted for publication in \emph{Geophysical Journal International} following peer review. The version of record is available at \url{https://doi.org/10.1093/gji/ggac151} and this work should be cited as:\vspace{0.4em}\\
Sambridge, M., Jackson, A. and Valentine, A.P., 2022. Geophysical Inversion and Optimal Transport. \emph{Geophysical Journal International}, ggac151, doi:10.1093/gji/ggac151.}}\end{center}

\begin{abstract}
We propose a new approach to measuring the agreement between two oscillatory time series, such as seismic waveforms, and demonstrate that it can be employed effectively in inverse problems. Our approach is based on Optimal Transport theory and the Wasserstein distance, with a novel transformation of the time series to ensure that necessary normalisation and positivity conditions are met. Our measure is differentiable, and can readily be employed within an optimization framework. We demonstrate performance with a variety of synthetic examples, including seismic source inversion, and observe substantially better convergence properties than achieved with conventional $L_2$ misfits. We also briefly discuss the relationship between Optimal Transport and Bayesian inference.
\end{abstract}

\section{Introduction}

In seismological studies---at global, regional and exploration scales---we often set out to find the earth structure and/or seismic source parameters that can explain observed seismic waveforms. In order to do so, we must first introduce some measure of the similarity between two waveforms, and hence define what it means for a synthetic seismogram to `agree with' observed data. The choice we make here can have far-reaching consequences, influencing not only the characteristics of the solution we obtain, but also the efficiency of the algorithms by which we obtain it.

By far the most commonly-used measure of similarity (or `misfit function') is the squared difference between the two waveforms, integrated over time: in other words, the square of the Euclidean (i.e., $L_2$) distance between the two. This is conceptually straightforward, mirroring our everyday notion of `distance'. Moreover, in cases where the waveforms are linear in the parameters sought (or may be approximated as such), this choice enables a highly-efficient solution strategy---the well-known `least-squares algorithm'. However, in many cases the relationship between Earth parameters and seismic waveform amplitudes is known to be non-linear; perhaps highly so. As a result, the Euclidean misfit can become complex with multiple minima, leading to difficult optimization problems. 

\begin{figure}
\begin{center}
\includegraphics[width=0.7\textwidth]{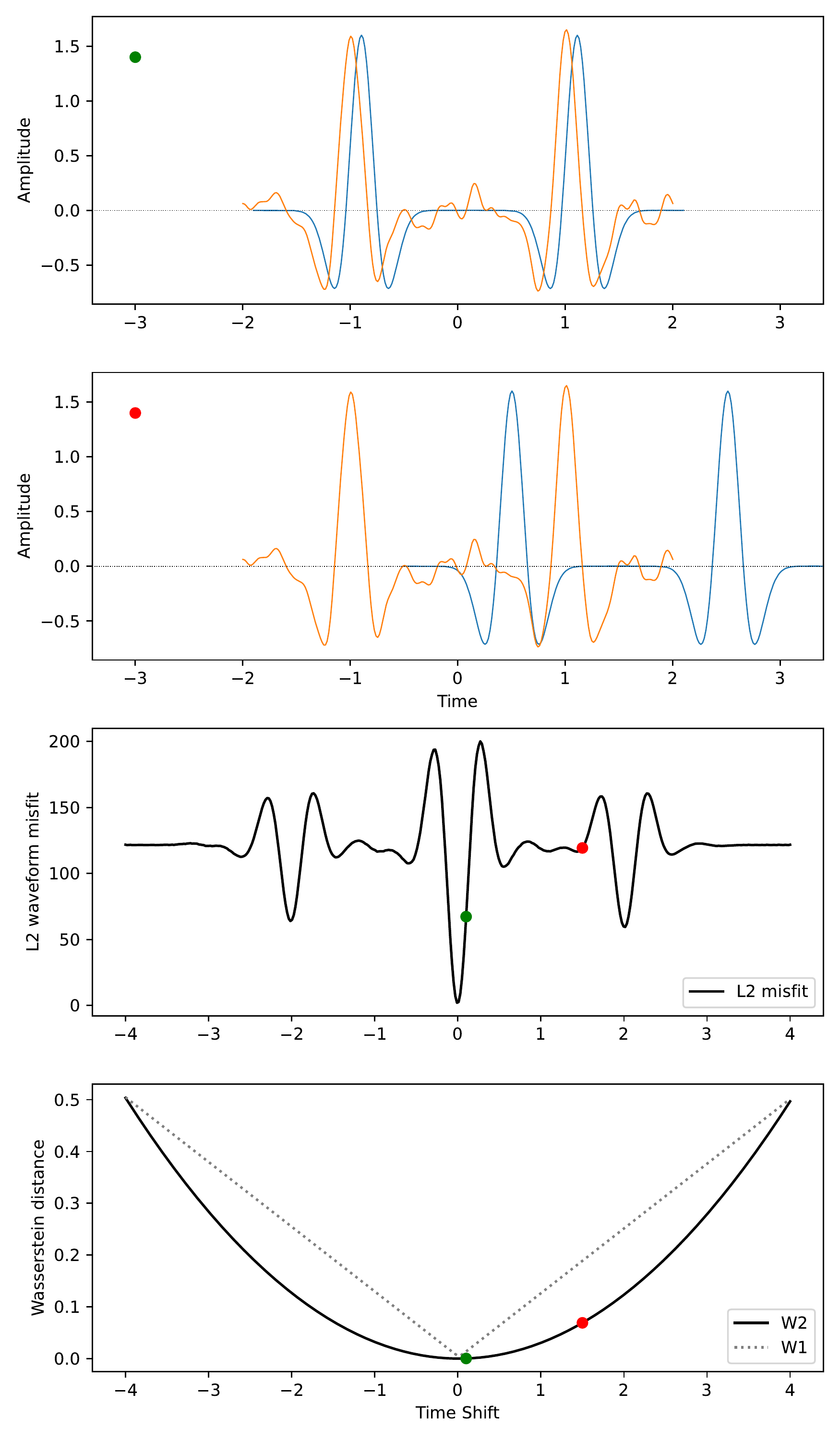}
\end{center}
\caption{\label{fig:ricker_align} Two scenarios for alignments of double Ricker wavelets. In the top panel the blue predicted wavelet is within half of one period of the noisy orange target so that local minimization of a squared waveform misfit for the time offset results in global alignment. Here  the orange curve has Gaussian noise added with correlation length 0.03s and amplitude of $5\%$ of the maximum signal height.
In the second panel the mis-alignment is sufficiently large to cause convergence to a secondary minimum, where the first peak of the blue is closer to the second peak of the orange. The bottom two panels show the squared waveform, $L_2^2$ misfit and two types of Wasserstein distance, $W_1$ and $W_2^2$ (see text), as a function of offset. The time offsets of the waveforms in the top panels are indicated by coloured dots on the lower two panels. The $L_2^2$ waveform misfit has many local minima while both the Wasserstein distances have a single global minimum.}
\end{figure}

One pragmatic approach may be to consider only small perturbations in parameters, allowing problems to be `linearized', and almost all seismic waveform inversion is to some extent based on linearization. However, a key issue is whether one has sufficient information about the unknown Earth or source parameters to allow an initial guess that is close enough to the best fit solution to render linearization useful in practice. In particular, for gradient-based optimization to converge, the predicted waveforms must be closely-aligned with the corresponding phase of the observed waveforms. This is illustrated in  Fig.~\ref{fig:ricker_align}: in the top panel two double Ricker wavelets are sufficiently close that minimization of $L_2^2$ will likely result in an optimal alignment, whereas in the second panel the first peak on the blue waveform is closer to the second peak of the orange waveform, which suggests that the same approach might result in the incorrect alignment of peaks. 

Various strategies have been developed to mitigate potential alignment issues, typically through preprocessing steps. Examples include first fitting arrival times of phases in body wave studies, or fitting seismic envelopes in surface wave studies. In reflection seismology it is usually necessary to first obtain the correct long wavelength velocity structure of a region, e.g. by travel time tomography, in order to align phases, as a prelude to full waveform inversion \cite[see e.g.][]{sirgue:2004}.  Without such alignment, gradient based methods that minimize the $L_2$ norm of the waveform difference can fail to converge, or converge to local minima in the misfit function. In exploration seismology this is commonly known as cycle skipping, the solution of which has been a long-standing area of research with many innovative approaches proposed, typically involving alternative strategies for measuring waveform misfit \cite[see e.g.][]{Luo:1991}, or multiple fitting procedures using different criteria. For a discussion of these issues in the context for full waveform inversion in exploration seismology see \cite{Gauthier:1986,Virieux:2009}, and for a recent review of proposed solutions see \cite{Pladys:2021}.

In this paper we explore a different measure of waveform similarity, based on Optimal Transport \citep{Villani:2003,Villani:2008}, which we believe can reduce or eliminate the need for such \emph{ad hoc} approaches. The theory of Optimal Transport can be traced back to the work of \citet{Monge:1781}, who sought to understand the most efficient way to distribute raw materials from sources to sinks; over the past two decades, it has undergone a renaissance and received much attention in the mathematics literature \citep[e.g.][]{Ambrosio:2003,Santambrogio:2015}. It was first introduced into geophysics by \citet{Engquist:2014} who pioneered its use in exploration seismology. Their work demonstrated its efficacy as a seismic misfit function for full waveform inversion, addressing the cycle skipping problem for high-frequency reflection waveforms. Optimal Transport-based misfit functions were found to increase the effective convexity of the objective function and hence the range of time shifts over which gradient-based algorithms would converge  \citep{Engquist:2014, Engquist_etal:2016, Yang:2018,Yang_etal:2018}. These results encouraged other authors to adopt the idea and develop variations on the theme in applications to full waveform imaging \citep{ Metivier:2016c, Metivier:2016a,Metivier:2016b,Metivier:2016d}. Although methods and details vary between studies, these contributions have now  established Optimal Transport as a powerful new approach for the construction of novel seismic waveform misfit functions in reflection seismology. More recently, the application of Optimal Transport has been explored for other geophysical problems including seismic receiver function inversion \citep{Hedjazian:2019}, gravity inversion \citep{Huang:2019}, and simulation of broad-band ground motions \citep{Okazaki:2021}. A number of useful surveys of the mathematical and computational aspects of Optimal Transport have been published, including those of \citet{Ambrosio:2003}, \citet{Santambrogio:2015}, \citet{Kolouri:2017}, \citet{Levy:2018} and \citet{Peyre:2019}. 

Inspired by this work, the present paper proposes a fairly general framework for seismic waveform fitting using Optimal Transport, which we believe is particularly suited to regional or global  settings, where observed and predicted waveform windows may be significantly mis-aligned in time and have large differences in amplitude. The approach may also have applications to other time series fitting problems that arise in the geosciences. We depart from previous methods developed for reflection seismology in both how Optimal Transport is applied to measure waveform discrepancy, and also the approach to solving the resulting Optimal Transport problem itself. Here the emphasis is on utilizing computationally efficient analytical solutions and calculating exact derivatives for use in nonlinear inversion.

In the following section we provide a brief introduction to Optimal Transport, particularly highlighting its application in 1D problems where convenient analytical solutions exist. The focus is on algorithms for exact calculation of Optimal Transport misfit functions, known as Wasserstein distances. In subsequent sections we present our new formulation for seismic waveform fitting together with our approach to Wasserstein distance calculation including its derivatives. In order to give the reader a concise description of our proposed algorithm, we focus on the main steps in the body of the paper, and leave many of the mathematical details to appendices.  We illustrate our approach through application to two waveform fitting problems. The first, motivated by earlier work, is a toy problem where the amplitude, time shift and dominant frequency of double-Ricker wavelets are solved for by minimizing different Wasserstein distances between a noisy and noiseless signal. The second is an example drawn from coupled seismic source location and moment tensor inversion of high rate GPS displacement waveforms, previously studied by 
\citet{OToole2011,OToole2012}.

\section{Optimal Transport}\label{sec:theory}

The original formulation of Optimal Transport by \citet{Monge:1781} focussed on finding efficient strategies for distributing resources (for example, coal) from sources (i.e. mines) to consumers. We introduce one 1D density function, $f(x)$, to represent the initial distribution, and another, $g(y)$, to represent the distribution of consumers---also known as the \emph{target distribution}. We assume that the overall volume of production is matched to consumption, and hence (with an appropriate choice of units) we can write
\begin{equation}
\int_{-\infty}^\infty f(x) \,\mathrm{d}x =\int_{-\infty}^\infty g(y) \,\mathrm{d}y = 1.
\label{eq:mass}
\end{equation}
Monge assumed that all the material initially in an interval $\delta x$  about point $x$ would be transported to some interval $\delta y$ about another point $y=T(x)$, with $f,g,T$ all $\ge 0$. Moreover, he assumed that this \emph{Transport Map}, $T(x)$, is one-to-one, i.e. no splitting of mass can occur. If this is to satisfy the needs of consumers, the Transport Map needs to be chosen such that
\begin{equation}
f(x) = g(T(x)) | \nabla T(x)|.
\label{eq:T}
\end{equation}
This expression represents the conservation of mass in transforming one density, $f(x)$ to another, $g(y)$. In general, there may be many maps satisfying this constraint. To find the \emph{optimal} strategy, we need to determine the cost associated with any particular choice. We introduce $c(x,y)$ to represent the cost of transporting a unit volume of material from point $x$ to point $y$; thus, the overall cost associated with distributing resources according to map $T$ is given by
\begin{equation}
I[T] = \int_{-\infty}^\infty c(x,T(x))f(x)\,\mathrm{d}x.
\label{eq:I}
\end{equation}
Monge focused on the case where the cost function has general form $c(x,y)=|x-y|^p$, and sought the optimal Transport Map that minimised the total cost functional, $I[T]$.

Subsequent work has proven that for the choice $p=2$, there is always a unique optimal Transport Map provided $f$ and $g$ are absolutely continuous with respect to the Lebesgue measure \citep{Brenier:1991}. In general, continuous formulations of this kind require difficult non-linear functional optimization. Several methods for the solution of Optimal Transport problems exist, depending on the choice of the distance function $c(x,y)$ and whether the density functions $f$ and $g$ are restricted to 1D (as in our description here), or 2D (as required for many practical problems). One class of approach is via the  numerical solution of the Monge-Ampere partial differential equation \citep{philippis:2013,Benamou:2000}. This also corresponds to $p=2$ and was used extensively in the earliest studies in geophysics \citep{Engquist:2014, Engquist_etal:2016, Yang:2018,Yang_etal:2018}.

\begin{figure}
\begin{center}
\includegraphics[width=0.9\textwidth]{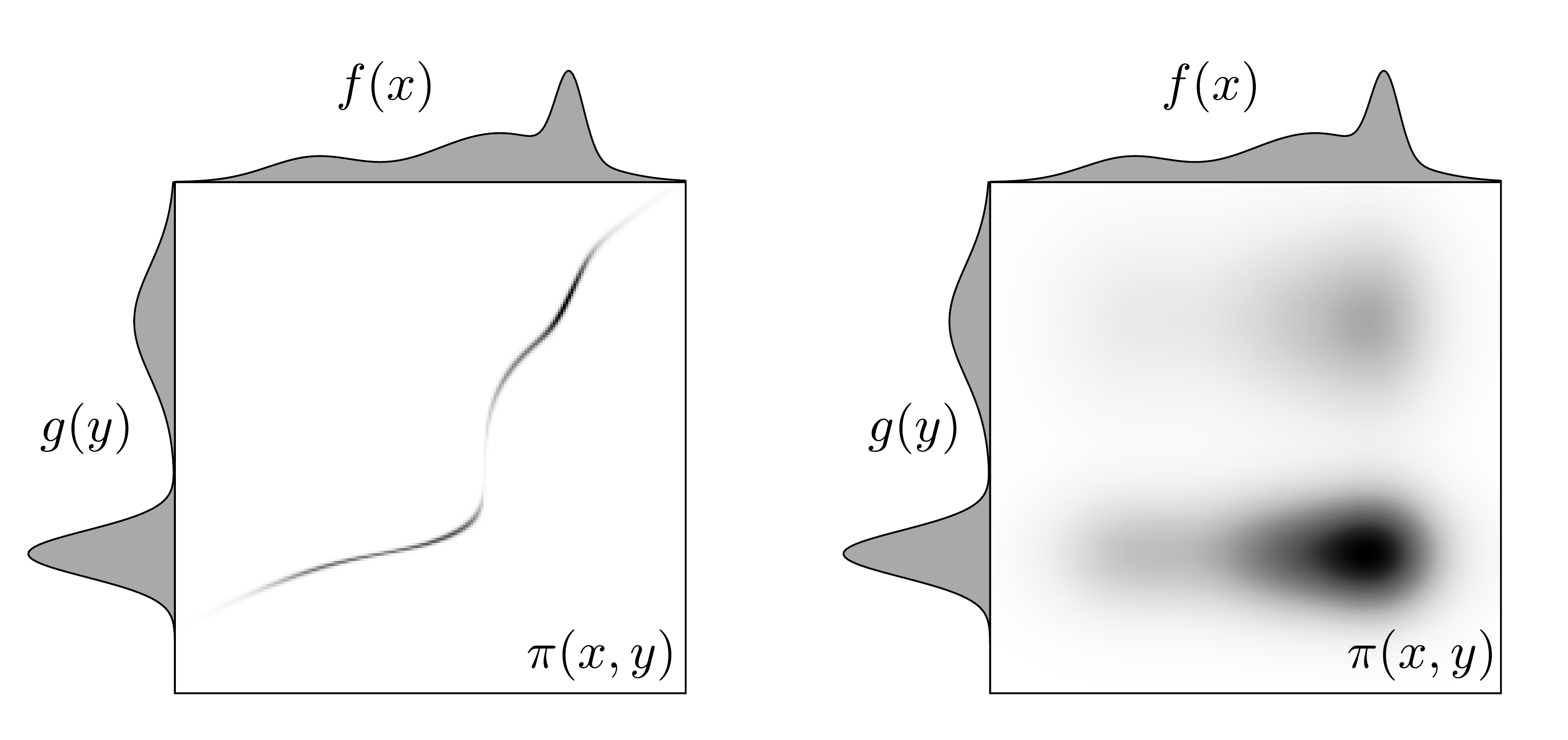}
\end{center}
\caption{\label{fig:tplans} \small Two examples of transport plans, $\pi(x,y)$, that have the same 1D marginal densities, $f(x)$ and, $g(y)$. The transport plan redistributes the mass from the source $f(x)$ to the target $g(y)$. In the left panel the 2D density, $\pi(x,y)$, is almost a single line, corresponding to a transport map, while on the right, mass is distributed from each point in $x$ to multiple regions along the $y$ axis, corresponding to a transport plan.}
\end{figure}

Following Monge's initial formulation, \cite{Kantorovich:1942} posed a more general problem by allowing mass to be split and combined during transport. Thus, Kantorovich's work is based upon the concept of a {\sl transport plan}, $\pi(x,y)$, chosen such that
\begin{subequations}\label{eq:tmarg}
\begin{align}
f(x) &= \int \pi(x,y)\,\mathrm{d}y\\
g(y) &= \int \pi(x,y)\,\mathrm{d}x
\end{align}
\end{subequations}
Again---as illustrated in Fig.~\ref{fig:tplans}---there are may be multiple plans that satisfy these requirements. The transport plan can be thought of as a recipe for redistribution of mass from $f(x)$ onto $g(y)$, where an infinitesimal region $\delta x$ at $x$ gets spread along the $y$ axis according to the corresponding conditional distribution, $\pi(y|x)$. The overall cost of the plan is given by
\begin{equation}
I[\pi] = \int c(x,y)\pi(x,y)\,\mathrm{d}x\mathrm{d}y
\end{equation}
and again one may seek the plan with minimal cost.

If $f(x)$ and $g(y)$ are functions of continuous variables $(x,y)$, and we henceforth refer to this as the continuous case, then it can be shown that the optimal plan, $\pi(x,y)$ i.e. $\displaystyle\min_{\pi} I(\pi)$, becomes a one to one map, for all $c(x,y)$ between $x$ and $y$. This is illustrated in the left-hand panel of Fig. \ref{fig:tplans} and this corresponds to Monge's Transport Map, $T(x)$. Hence they each lead to the same result. For the discrete case \cite{Kantorovich:1942} showed that, for any $p$, the problem of finding $\pi$ reduces to one of Linear Programming, details of which appear in Appendix \ref{app:LP}. This was important because the existence of a solution in Monge's nonlinear approach, (\ref{eq:I}), is not guaranteed, whereas a solution can always be found with  Kantorovich's Linear Programming (LP) formulation.   This approach is also completely general, in that it can find solutions for any $p$ value with $f$ and $g$ in any number of dimensions. The disadvantage of the LP formulation is that it becomes computationally prohibitive as the number of discrete elements in $f$ and $g$ increases, not least because the number of unknowns to be solved for in the transport matrix, $\pi_{ij}$, is equal to the product of the number of elements in $f$ and $g$. The computational cost can be reduced somewhat with primal-dual optimization methods. For $p=1$ this is equivalent to minimization of the Kantorovich-Rubenstein norm between density functions \citep{Lellmann:2014}, which has also been extensively applied in seismic full waveform inversion of reflection data \citep{Metivier:2016a,Metivier:2016b, Metivier:2016c, Metivier:2016d} and more recently to receiver function inversion \citep{Hedjazian:2019} as well as body and surface waves in land seismics \citep{He:2019}. Another method for 2D discrete cases with $p=2$ involves using a regularized Linear Programming formulation through addition of an entropy term. This introduces an element of smoothing to the transport plan but allows more efficient algorithms \citep{Cuturi:2013,Solomon:2015,Solomon:2015a}, and is popular within the computer science literature. 

Overall, we see a variety of different approaches to solving Optimal Transport problems have been developed, reflecting different choices regarding the value of $p$, the dimension of the density functions, and whether the problems are continuous or discrete. 
In this paper our formulation of the problem for time series means that we only need to solve Optimal Transport problems in 1D. We now briefly describe some results for the 1D case which we will use later as the basis of an efficient discrete solution algorithm.

\subsection{Analytical solutions in 1D}\label{sec:1Dsol}
When $f(x)$ and $g(y)$ represent 1D densities, and $x$ and $y$ are real variables, a unique Transport Map solution exists for all distance functions of the form 
$c(x,y) = |x-y|^p$, for $p> 1$
\citep{Villani:2003,Villani:2008}. Typically this is expressed through the $p$-Wasserstein distance, which we have loosely referred to above, and  may now be defined from (\ref{eq:I}) as
\begin{equation}
W_p(f,g) = \left[\inf_{T\in\mathcal{T}} \int c(x,T(x)) f(x)\,\mathrm{d}x\right]^{1/p},
\label{eq:W}
\end{equation}
where $\mathcal{T}$ represents the space of all transport plans satisfying (\ref{eq:T}).
The Wasserstein distance, $W_p(f,g)$, is a measure of the work done in redistributing $f$ to $g$. Substituting in the general cost functions we have
\begin{equation}
W_p(f,g) = \left[ \int_X |x-T_\star(x)|^p f(x)\,\mathrm{d}x\right]^{1/p},
\label{eq:W2}
\end{equation}
where we use $T_\star(x)$ to denote the optimal Transport Map.  \citet{Villani:2003} has shown that for the 1D case there exists a simple analytical solution for $W_p$ with two alternative forms
\begin{equation}
W_p^p(f,g) =\int_X |x-G^{-1}(F(x))|^p f(x)\,\mathrm{d}x 
\label{eq:sol0},
\end{equation}
or equivalently,
\begin{equation}
W_p^p(f,g)  = \int_0^1 |F^{-1}-G^{-1}|^p \,\mathrm{d}t,
\label{eq:solW2}
\end{equation}
where $F(x)$ and $G(y)$ are the cumulative distribution functions of $f(x)$ and $g(y)$ respectively
\begin{equation}
F(x)  = \int_{-\infty}^x f(x^{\prime}) \,\mathrm{d} x^{\prime}, \quad G(y)  = \int_{-\infty}^y g(y^{\prime}) \,\mathrm{d} y^{\prime},
\end{equation}
and $F^{-1}(t)$ and $G^{-1}(t)$ are the inverse cumulative distribution functions of the variable $t$, where $ 0 \le t \le 1$.

The expression on the right hand side of (\ref{eq:solW2}) is the integral of the absolute difference between the inverse cumulative distributions to the power of $p$.
For $p=1$ the integral (\ref{eq:solW2}) corresponds to the area between the two curves, and since the integration may be performed along either axis then we also have an equivalent expression
\begin{equation}
W_1(f,g) =\int_X |F-G| \,\mathrm{d}x,
\label{eq:intcum}
 \end{equation}
which avoids the calculation of the inverse cumulative distributions. Furthermore by comparing (\ref{eq:W2}) and (\ref{eq:sol0}) one can see that 
for the 1D continuous case the Transport map is given under any $p$-norm by
\begin{equation}
T_\star(x) = G^{-1}(F(x)).
\label{eq:Tmap}
 \end{equation}

\section{Application to time series}\label{sec:alg}

We will shortly discuss how the preceding results may be implemented as numerical algorithms. However, it is convenient to first discuss the problem of applying Optimal Transport to the comparison of two arbitrary time series.
As the discussion above shows, Optimal Transport is conceived as a method for transforming between density functions, which are always positive and integrate to one. Time series are generally oscillatory, with both positive and negative values; they may or may not integrate to any known value. Some scheme is therefore required to transform time series into a format amenable to application of Optimal Transport theory, and previous authors have adopted a variety of strategies. This is a key issue in the effectiveness of any approach, and \citet{Mainini:2012} critically evaluates the merits of various transforms from time series to density-like functions. The simplest option is to add an arbitrary scalar to the amplitudes of all waveforms to ensure positivity. However, this is unsuitable for an inversion application, since the amplitude of the predicted seismic waveforms depend on unknown Earth model parameters and cannot be known in advance.

Two other popular choices, both  considered by \citet{Engquist:2014}, are firstly taking the absolute value of waveforms; and secondly, the separate transport of positive to positive and negative to negative components between pairs of waveforms. Drawbacks of the first approach include a loss of polarity information in the waveform, and a lack of differentiability at times where the time series crosses the zero axis, both of which may be detrimental in an inversion context. In the second transform two Wasserstein distances are calculated, i.e. one between positive parts of observed and predicted waveforms and a second between negative parts of the waveforms. This has the undesirable quality of creating an artificial decorrelation between positive and negative parts of the signal. It also does not {\sl preserve mass}, i.e. integrals of the waveform over the time window \citep{Mainini:2012}. 

Of the other transforms reviewed by \cite{Mainini:2012} the most popular seems to be the {\sl global strategy} that performs transport between two composite signals, constructed by combining the positive part of each waveform with the (inverted) negative part of the other. This is popular among some authors because it is both symmetric and preserves mass, but it has the disadvantage (in the context of waveform fitting for inversion) that it blurs the distinction between observed and predicted waveforms. In some circumstances---such as if the original waveforms happen to be separated significantly in time---it is possible that transport will occur between the positive and negative components of the same signal, i.e. the Wasserstein distance of misfit might reflect the transport of the positive parts of the predicted waveform to the negative parts of the predicted waveform rather than the observed waveform. It can be shown that this global strategy underlies the primal-dual optimization ($p=1$) approach employed in the full waveform inversion studies of \cite{Metivier:2016a,Metivier:2016b, Metivier:2016c, Metivier:2016d}. In another, more recent contribution, \citet{Sun:2019} have proposed using Optimal Transport to compare a matching filter, calculated from the deconvolution of a predicted and observed waveform, with a representation of the Dirac delta function. In this case the objects being compared are both naturally positive and hence lend themselves to application of Optimal Transport.

It is clear that there is no universally accepted procedure for imposing positivity on time series, or more generally creating density functions for use with Optimal Transport. Different studies use differing approaches and each have drawbacks. In a departure from earlier trends,  \citet{Metivier:2018,Metivier:2019}  proposed a novel idea called `Graph space OT', or GSOT. In their study they assign a series of delta functions along both the observed and predicted waveforms and perform transport between the 2D density functions formed by their sum in the time-amplitude plane.  In this way the positivity issue is circumvented by recasting the waveform transport problem as one of integer assignment between two point clouds in 2D, which can be solved with the `Auction' algorithm of \cite{Bertsekas:1989}. Projecting the problem into 2D is appealing because the positive or negative amplitudes of the original waveforms are treated equally and merely determine  the location of points in time-amplitude space. It does, however, depend on choices made in defining a Euclidean distance in time-amplitude space. This approach continues to be applied to an increasing range of full waveform inversion problems \citep[see][]{Gorszczyk:2021}.

In our approach below, we also recast the transport problem from 1D to 2D, but rather than using a discrete point cloud we build a 2D density function across the time-amplitude plane and use this as the basis of Wasserstein calculations. As will be seen, the shape of our 2D density function extends the character of a waveform across the time-amplitude plane without the need to treat positive and negative amplitudes separately. For computational convenience we project our 2D density function transport problem back to 1D through marginalisation, which allows use of the exact Optimal Transport solutions given in section {\ref{sec:theory}. To aid clarity, we describe our step-by-step procedure below, with some mathematical details confined to the appendices. When these steps are combined, they form a procedure for calculating the Wasserstein distances between an observed waveform, $u_{obs}(t)$, and its synthetic counterpart, $u_{pre}(t)$, which exist in independent time-amplitude windows that may, or may not, overlap. Importantly, each step is designed to be a differentiable transform of the waveforms, so that application of the chain rule allows evaluation of the gradient of the Wasserstein distance with respect to model parameters---as is required if the misfit function is to be the basis of any gradient-based optimisation procedure.

\subsection{Time-Amplitude windows}\label{sec:tuwin}

The first step of our approach is to define a finite time-amplitude window for both the observed and predicted waveforms, as shown in Fig. \ref{fig:rickerfp}. 
We assume that independent time windows, $t \in (T_0, T_1)$ and $t \in (T_2, T_3)$, are constructed for observed and predicted waveforms respectively, which may or may not be overlapping. In this case a simple affine transform in time will suffice to produce a non-dimensional time parameter $t^{\prime}$.  Specifically we have for the observation window
\begin{equation}
t^{\prime} = \frac{t-T_0}{\Delta_t},
\label{eq:affine}
\end{equation}
where $\Delta_t = T_1 - T_0$, and clearly $t^{\prime}  \in [0,1]$ for the observation window, and  $t^{\prime}  \in [\frac{T_2-T_0}{\Delta_t},\frac{T_3-T_0}{\Delta_t}]$ for the predicted window. Similarly, it is straightforward to define an amplitude window that is sufficient to encompass the observed waveform, $\Delta_u = u_1 - u_0$. However, as previously discussed, it cannot be guaranteed that the predicted waveform will also lie in this amplitude range, as this will depend on the model parameters chosen. We therefore wish to apply a transformation that maps the entire amplitude axis, $[-\infty,+\infty]$, to a finite range, i.e. $u^{\prime} \in [0,1]$, while retaining maximum sensitivity within the interval $[u_0,u_1]$. A convenient choice is based on the inverse tangent function,
\begin{equation}
u^{\prime} = \frac{1}{2} + \frac{1}{\pi} \tan^{-1} \bar u
\label{eq:utrans0}
\end{equation}
where
\begin{equation}
\bar u = \frac{1}{\Delta_u}[2u - u_0 - u_1].
\label{eq:utrans1}
\end{equation}
In practice we might choose to extend the amplitude parameters $(u_0,u_1)$  beyond the amplitude range of the observed waveform, $u \in [a_{min},a_{max}]$,  by about $20\%$, e.g. $u_0 = a_{min} - 0.1\Delta_a, u_1 = a_{max} + 0.1\Delta_a$, where $\Delta_a= a_{max}-a_{min}$.  Combining these two transforms produces the desired outcome of a pair of time-amplitude windows, with a common amplitude range, guaranteed to contain both observed and predicted waveforms, and with maximum sensitivity about the desired interval of the observed waveform.

In most seismic inversion applications, derivatives of waveform amplitudes with respect to Earth model, or seismic source, parameters will be required and we assume these are available from a given forward model. Since (\ref{eq:utrans0})-(\ref{eq:utrans1}) transform the original waveform amplitudes, $u$, to non-dimensional amplitudes, $u^{\prime}$, then we will require the derivative of the transformed amplitude with respect to the untransformed amplitude. This is straightforward and given by 
\begin{equation}
\frac{\partial u^{\prime}}{\partial u} = \frac{2}{\pi \Delta_u} \frac{1}{(1+\bar u^2)}.
\label{eq:utrans0d}
\end{equation}

\begin{figure}
\begin{center}
\includegraphics[width=0.7\textwidth]{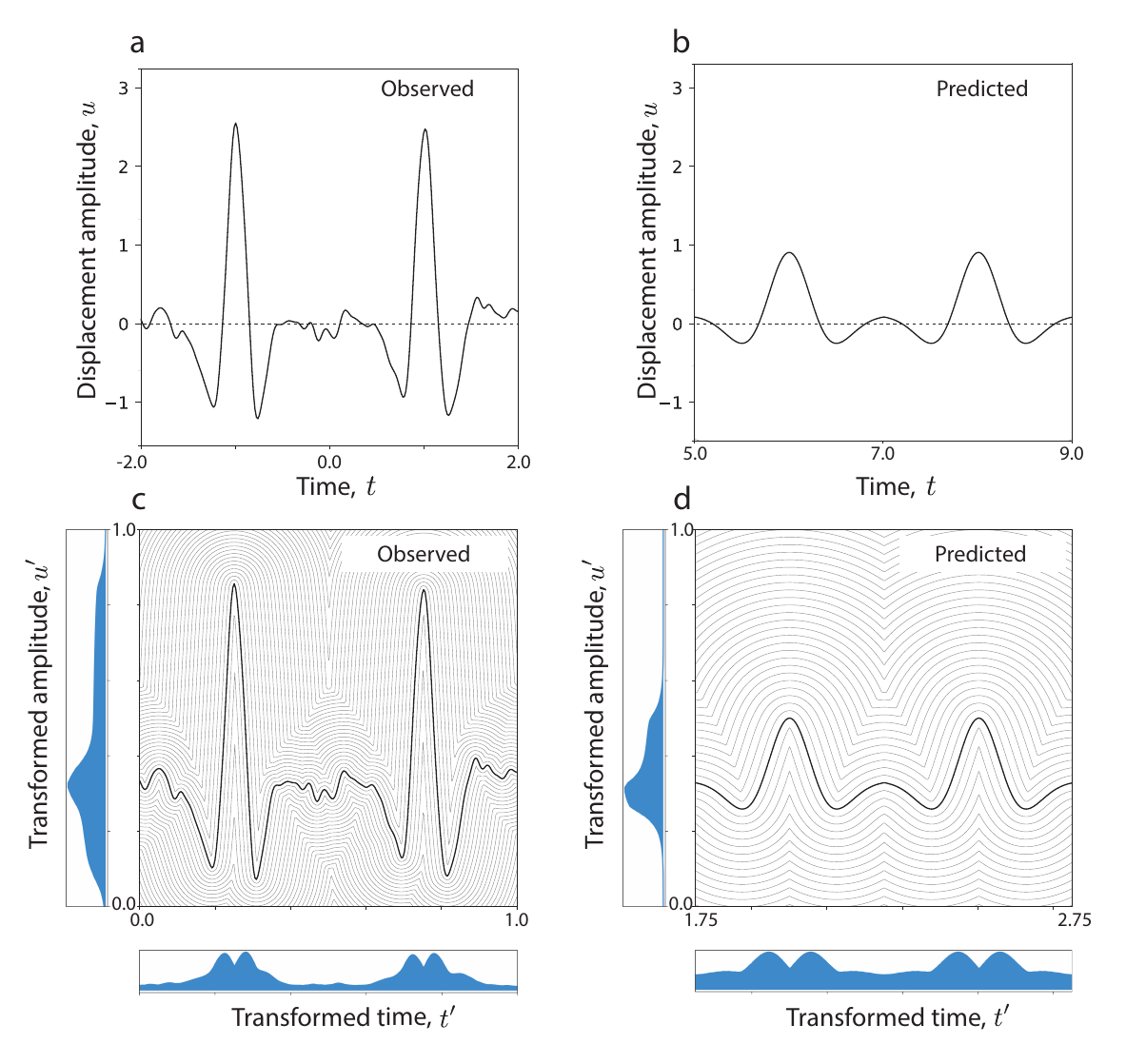}
\end{center}
\caption{\label{fig:rickerfp} a) \& b) show two example double Ricker wavelets, $R_1(t)$ and $R_2(t)$, respectively, with differing dominant frequency and amplitudes, and windows displaced in time by 7s. The wavelet, $R_1(t)$, has Gaussian noise added with correlation length 0.03s and amplitude of $5\%$ of the maximum signal height. For each window we have $\Delta_t=4s$, and the Time-Amplitude transform (\ref{eq:affine}) - (\ref{eq:utrans1}) produces non-dimensional variables $(u^{\prime},t^{\prime})$ and a $[0,1]\times[0,1]$ window for the observed waveform. c) \& d) show 2D nearest distance fields  calculated for each curve  as described in the main text. These have the appearance of {\sl fingerprints} and represent the information content of the waveform extended across the time-amplitude plane. The shaded density functions on the amplitude and time axis for panels c) \& d) are 1D marginals, derived from 2-D density fields described in section \ref{sec:algPDF}, which are used to represent misfit between the two curves. In the experiments of section  \ref{sec:drickerfit} the best time shift, amplitude and frequency parameters are found by minimizing the Wasserstein distances between  `observed' and `predicted' marginals.}
\end{figure}

\subsection{Seismogram fingerprints}\label{sec:algfinger}

The second step of our approach is to convert each 1D transformed waveform, $u^{\prime}(t^{\prime})$, into a representative 2D function, over its respective transformed time-amplitude window.  A convenient choice is the distance from each point within the time-amplitude window to the nearest point on the waveform, $u^{\prime}(t^{\prime})$. In practice this calculation is discretised onto a regular $(n_t \times n_u)$ grid and represented as $d_{ij}, (i=1,\dots,n_t;j=1,\dots,n_u)$. Figs. \ref{fig:rickerfp}c \& d  and \ref{fig:FPandPDF} show examples of a contoured nearest distance field for double Ricker wavelets and a seismic receiver function, respectively.  As can be seen, this has the appearance of a `fingerprint' that contains within it the original waveforms, as the zero distance contour, but spreads the influence of each amplitude variation across the entire plane. The contours of the 2D nearest distance field contain the full information content of the oscillatory waveform in a higher dimensional object, and avoid the need to specifically manipulate negative parts of the waveform. This idea is motivated by the work of \citet{SethianA:1999} who found that using nearest distance fields constructed from handwriting examples, helped improve the success of optical character recognition algorithms. In our case, it is appealing because the information content previously confined to a complex 1D object (a waveform) has been converted to a  2D image. 
Since the distance field, $d_{ij}$, is calculated in the transformed time-amplitude plane, then clearly the relative influence of amplitude versus time variations on the shape of its contours (Fig. \ref{fig:rickerfp}c \& d) will depend on the choice of the ratio $\Delta_u/\Delta_t$. 

It transpires that to find the nearest distance field, $d_{ij}$, it is not necessary to exhaustively search over the complete discretized waveform for the minimum distance to each grid point, as there are several more efficient methods available. Indeed this calculation may be familiar to a seismologist because it is identical to that of constructing a seismic travel time field with geometric ray theory. In particular, precisely the same distance contours are obtained if we consider the receiver function in Fig. \ref{fig:FPandPDF} as a seismic wavefront at zero-time, and then let it propagate both in the positive and negative amplitude directions across the plane in a homogenous medium with unit velocity. Methods based on the Fast Marching Algorithm \citep{Sethian:1996,RawlinsonA:2002} are available for this task. In our study we represent the waveform as a series of linear segments, rather than discrete points, and so $d_{ij}$ is the distance from the $(i,j)$th grid point to the nearest point on the segmented wavefront. For this higher order representation Fast Marching is not ideal and the method we employ is described in Appendix \ref{app:FP}.
 For this second step we also require derivatives of $d_{ij}$ with respect to the non-dimensional amplitude of the waveform, i.e. $\frac{\partial d_{ij}}{\partial u_k^{\prime}}, (i=1,\dots,n_t; j=1,\dots,n_u; k=1,\dots,n)$, where $u_k^{\prime} = u^{\prime}(t_k^{\prime})$. Given the equivalence to geometric ray theory, it turns out that the distance $d_{ij}$ only depends on  the waveform amplitudes, $u_k^{\prime}$, of the segment containing the nearest point.  Hence a maximum of two derivatives are non-zero for each $(i,j)$. Furthermore, analytical  expressions for the derivatives are easily obtained since they are closely related to well known expressions for derivatives of travel times with respect to seismic source location in a homogeneous medium. Full details are given in Appendix  \ref{app:FP}.

\subsection{Density fields}\label{sec:algPDF}
The diligent reader will have spotted that the distance field $d_{ij}$ is zero on the original ray and increases away from it. Since our aim is to create a density function for use with Optimal Transport,  it seems reasonable to presume that the highest density should be on the waveform, and decrease away from it, rather than the reverse. The third step of our approach is then simply to convert the distance field to a density, by taking its negative exponential. Specifically we write the unnormalised density at the $(i,j)$ grid point, $p_{ij}$, as
\begin{equation}
p_{ij} = \exp\left(-d_{ij}/s\right),
\label{eq:pdfg}
 \end{equation}
where $s$ is an arbitrary scale factor. Together with the ratio $\Delta_u/\Delta_t$, the scale factor, $s$, determines the distance metric controlling the influence of the original waveform on the density amplitudes, $p_{ij}$. The lower panel of Fig. \ref{fig:FPandPDF} shows an example of the 2D density function formed from the receiver function in the top panel for the case $s=0.04$.
To satisfy the mass conservation conditions necessary for Optimal Transport (\ref{eq:mass}), we normalise the density amplitudes 
\begin{equation}
\bar p_{ij} = \frac{p_{ij}}{\sum_{q} \sum_{r} p_{qr}},
\label{eq:pnormg}
 \end{equation}
which yields a form suitable for input to a Wasserstein distance calculation.  Choices other than (\ref{eq:pdfg}) are of course possible, e.g. using the square of the distance or other function. We have not explored these possibilities, and merely content ourselves that the resulting 2D density function shown in Fig. \ref{fig:FPandPDF} looks to be reasonable, in that it has maximum weight on the original waveform and propagates the influence of the waveform shape across the time-amplitude plane. As in the earlier steps we need to keep track of derivatives, in this case of the normalised (or unnormalised) density function amplitudes with respect to the distance field, which means we require $\frac{\partial p_{ij}}{\partial d_{ij}}, (i=1,\dots,n_t; j=1,\dots,n_u)$ and $\frac{\partial \bar p_{ij}}{\partial p_{qr}}, (i,q=1,\dots,n_t; j,r=1,\dots,n_u)$. Again these may be obtained analytically in a straightforward fashion, and full details are given in Appendix \ref{app:FP}.

\begin{figure}
\begin{center}
\includegraphics[width=0.7\textwidth]{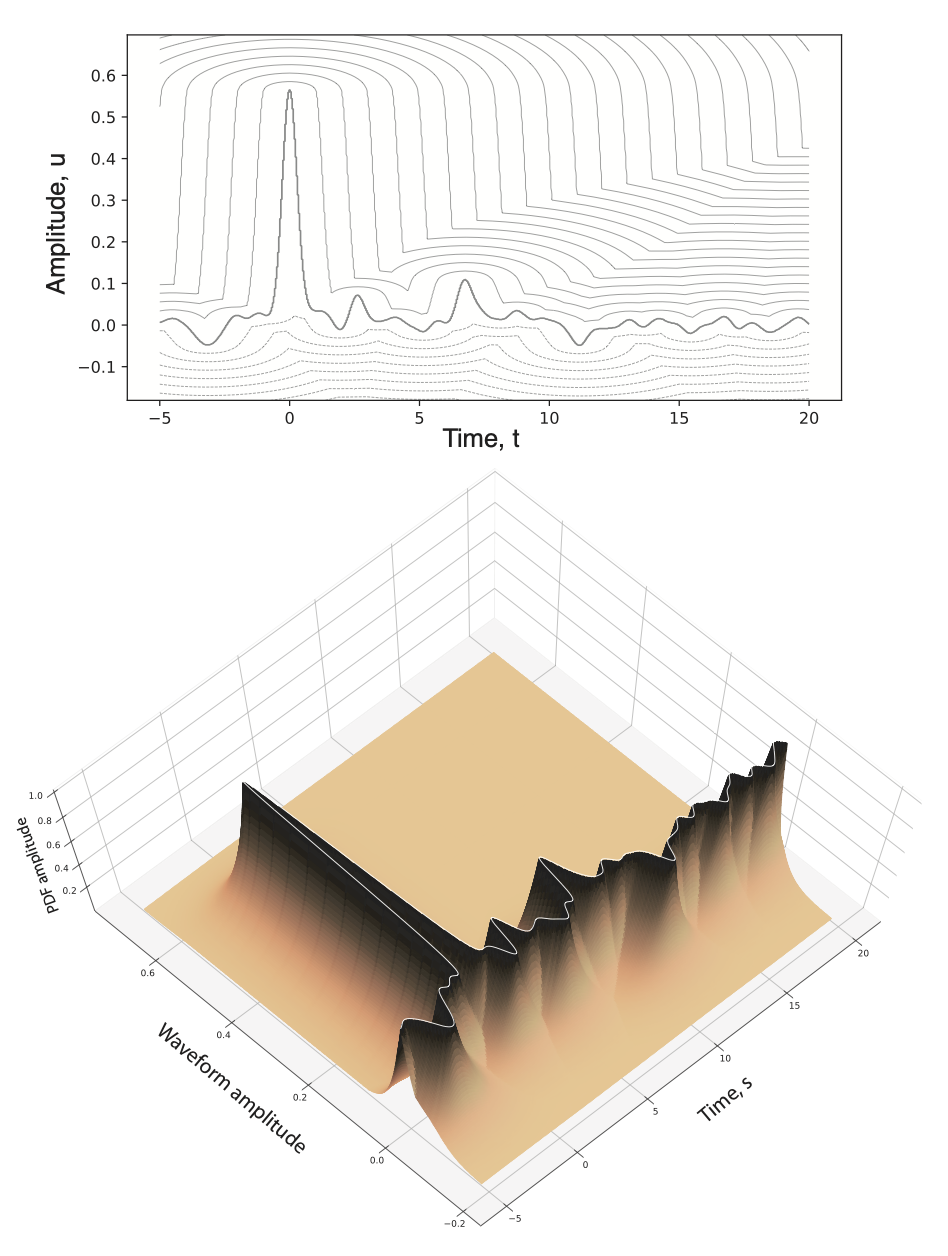}
\end{center}
\caption{\label{fig:FPandPDF} Top panel shows a seismic receiver function and its associated nearest distance field contours. The lower panel shows the 2D   density function created by the distance field using (\ref{eq:pdfg}) with a scale factor of $s = 0.04$. Larger values create a broader  function about the original waveform that forms a `ridge' with the highest density. 2D density functions can be constructed for both observed and predicted waveforms and the Wasserstein distance between them calculated using Optimal Transport, which becomes the misfit function in our waveform fitting algorithm.}
\end{figure}

\subsection{Marginal Wasserstein}\label{sec:algWmarg}
The final step of our procedure is to calculate Wasserstein distances between  observed and predicted 2D density functions, as in Fig. \ref{fig:FPandPDF}. While there are several methods available for 2D Wasserstein calculations, each involve the numerical solution of either a partial differential equation, as in the Monge-Ampere solvers, or numerical solution of an optimization problem, as for Linear Programming based solvers and primal-dual methods. Recently, much effort has been devoted to developing computationally and memory efficient 2D methods based on entropically-smoothed Wasserstein measures \citep{Cuturi:2013,Solomon:2015,Solomon:2015a}. However these involve the  Optimal Transport  between 2D density functions like that in Fig. \ref{fig:FPandPDF}, each of which is over a grid of $n_t \times n_u$ discrete elements. Typically the computational costs of many of these methods will scale with the cube of the size of the input, hence $\propto n_t^3 n_u^3$ in the worst case. This is unattractive for our purpose, where applications are likely to involve computing misfits between many pairs of waveforms. Rather than calculate the Wasserstein distance in 2D, our solution is to recast the problem as two 1D Optimal Transport problems by replacing the density function with its amplitude and time marginals. Specifically, we integrate the 2D density function in ({\ref{eq:pdfg}) along the amplitude axis to produce a 1D time marginal, $p^{(t)}$, and also along the time axis to produce a 1D amplitude marginal, $p^{(u)}$.  We have 
\begin{equation}
p^{(t)}_i = \sum_j^{n_u} \bar p_{ij}, \quad (i=1,\dots,n_t)
\label{eq:margdef1}
\end{equation}
and
\begin{equation}
p^{(u)}_j = \sum_i^{n_t} \bar p_{ij}, \quad (j=1,\dots,n_u).
\label{eq:margdef2}
 \end{equation}
Fig. \ref{fig:rickerfp}c \& d, shows examples of these marginals shaded along side each axis, for the double-Ricker wavelet case. The shape of each marginal is in general dependent on both time and amplitude variations in the original waveform.

In this way Wasserstein distance calculations between observed and predicted counterparts, which we write as $W_t(p^{(t)}_{obs}, p^{(t)}_{pre})$ and  $W_u(p^{(u)}_{obs}, p^{(u)}_{pre})$, require only the solution of 1D Optimal Transport problems. This significantly reduces the computational complexity of the task, and allows us to exploit the analytical solution given by (\ref{eq:solW2}). Our algorithm for evaluating 1D Wasserstein distances is described in the following section.

It is clear that some information must be lost through use of marginalization, but as we shall show, this does not appear to be a  disadvantage because in all cases considered here, we observe that alignment of the pair of 1D marginals uniquely corresponds to alignment of the original waveforms and importantly therefore no significant non-uniqueness is introduced in these cases. This suggests that 1D marginals provide an effective representation of the information content of the 2D density function.

Since our procedure results in two Wasserstein distances, we have the option to minimize their weighted sum with respect to Earth model parameters, which we write as
\begin{equation}
W^p_p(u_{obs},u_{pre}) = \alpha ~ ^tW^p_p\left(p^{(t)}_{obs}, p^{(t)}_{pre}\right) +  (1-\alpha)~^uW^p_p\left(p^{(u)}_{obs}, p^{(u)}_{pre}\right),
\label{eq:Wopt}
 \end{equation}
where the subscript in $W^p_p$ denotes the order-$p$ Wasserstein distance, the superscript indicates that it is to the power of $p$, the first superscript in  $^tW^p_p$ and  $^uW^p_p$ indicates either the time or amplitude marginal, and $\alpha$ reflects the choice of relative weighting, $0\le \alpha \le 1$. 
For inversion, two sets of derivatives are relevant here to connect to the previous step. Firstly, derivatives of the 1D marginal amplitudes with respect to the 2D density amplitudes, $(\partial p_i^{(t)}/ \partial p_{qr}, \partial p_j^{(u)}/ \partial p_{qr})$, which are given in appendix \ref{app:FP}, and secondly the derivatives of the Wasserstein distances with respect to the marginal amplitudes, $\partial ~^tW^p_p/\partial p_i^{(t)}$ and $\partial ~^uW_p^p/\partial p_j^{(u)}$. Calculation of Wasserstein distances between 1D density profiles is discussed in the next section, and the relevant derivatives in appendix \ref{app:margwass}.  Again, in all cases we are able to derive analytical expressions that may be evaluated efficiently and without approximation.

\subsection{An algorithm for Wasserstein distances between density profiles}\label{subsec:dpm}

\begin{figure}
\begin{center}
\includegraphics[width=0.9\textwidth]{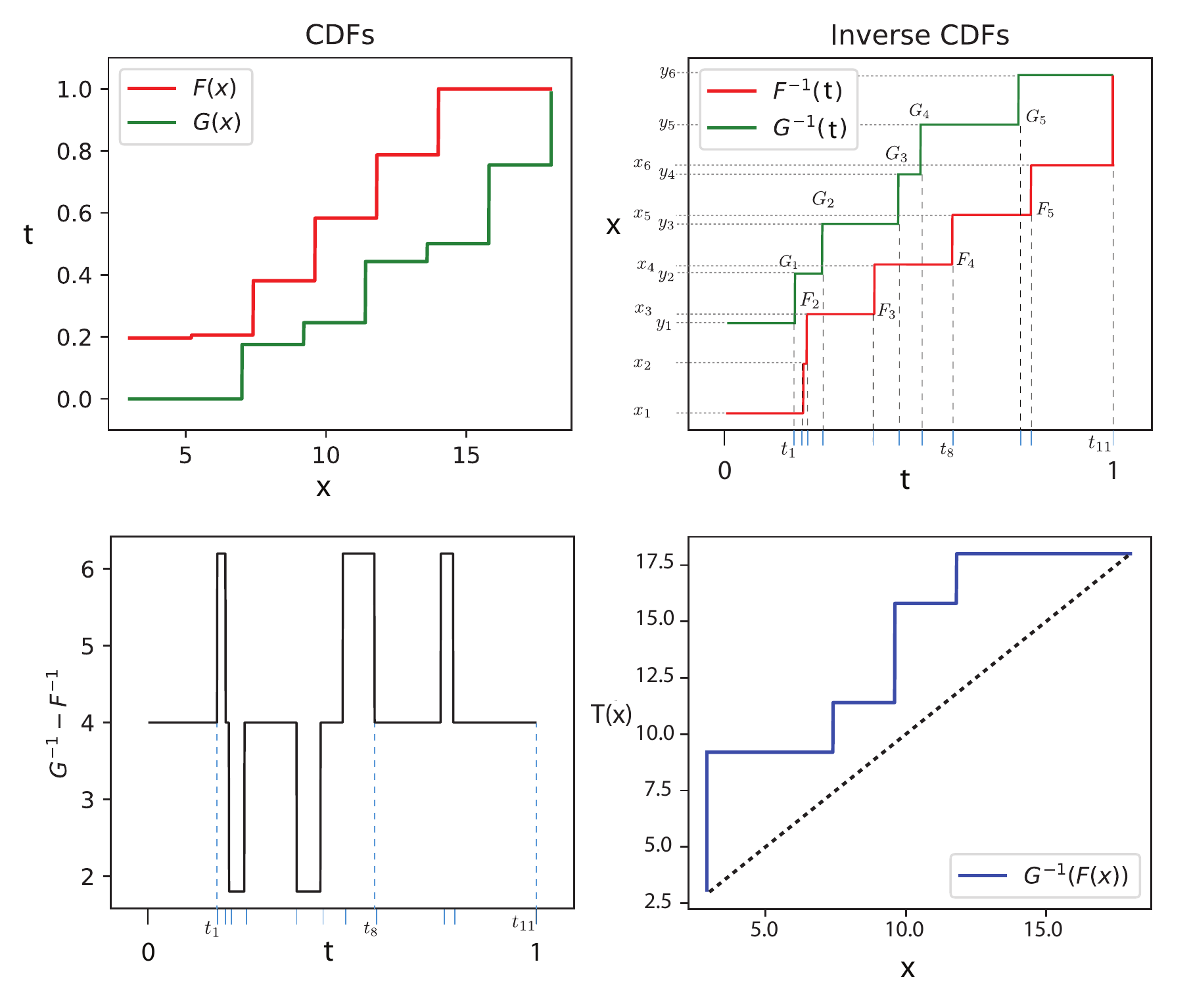}
\end{center}
\caption{\label{fig:1D} The top panels illustrate cumulative distribution functions (CDFs), together with their inverses, for two 1-D discrete densities, $f$ and $g$, each consisting of a sum of 6 delta functions at locations $x_i, (i=1,\dots, 6) $ and $y_j, (j=1\dots,6)$, respectively, where $x_i = 3+2.2(i-1)$ and
$y_i = 7+2.2(i-1)$. These correspond to eqns. (\ref{eq:pointmass}a \& b) with weights $(f_1,f_2,\dots) = (0.2,0.01,0.18,0.21,0.2,0.2)$ and $(g_1,g_2,\dots) = (0.18,0.07,0.2,0.05,0.27,0.23)$.  We make use of a variable $t$ to represent the range of the CDF, or equivalently the domain of the inverse CDF function. In the top right panel the change points in the piecewise constant inverse CDFs are indicated on the $t$-axis by the $t_k, (k=1,\dots, 10)$.  Note that $t_{11}$ is always unity. The lower panels show the difference between the inverse CDFs (left) as a function of $t$ and the optimal Transport Map between $f$ and $g$ (right).}
\end{figure}

In order to implement the final step of our approach we require an efficient and exact method for solving Optimal Transport problems in 1D.
Our algorithm for doing so is based on evaluation of (\ref{eq:solW2}) for the case where $f$ and $g$ may be represented as a  weighted sum of delta functions, i.e.
\begin{subequations}\label{eq:pointmass}
\begin{align}
f(x) &= \sum_{i=1}^{n_f} f_i \delta (x-x_i)\\
g(y)  &= \sum_{j=1}^{n_g} g_j \delta(y-y_j)
\end{align}
\end{subequations}
with the delta function $\delta(x-x_i)$ representing a unit point mass at position $x_i$. Here we consider the general case where $n_f \ne n_g$, although in our application, they will both be equal to either $n_t$ for the time marginal, or $n_u$, for the amplitude marginal.  Similarly we do not assume that the locations of the delta functions, ($\{x_i\,:\, i=1\ldots n_f\}$ and $\{y_j\,:\, j=1\ldots n_g\}$), coincide. To satisfy the normalisation constraint (\ref{eq:mass}), we require 
\begin{equation}
\sum_{i=1}^{n_f} f_i = \sum_{j=1}^{n_g} g_j = 1.
\end{equation}
These densities have cumulative distribution functions, which take the form of piecewise continuous `staircases',
\begin{subequations}
\begin{align}
F(x) &= \begin{cases}0&x_1 >  x\\F_i &x_i\le x< x_{i+1}, \quad (i=1,\dots,n_f-1)\\1&x_{n_f} \le x\end{cases}\label{eq:cumul1Df}\\
G(y) &= \begin{cases}0&y_1 > y\\G_j &y_j\le y< y_{j+1}, \quad (j=1,\dots,n_g-1)\\1&y_{n_g} \le y\end{cases}\label{eq:cumul1Dg}
\end{align}
\end{subequations}
where $F_i = \sum_{k=1}^i f_k$ and $G_j=\sum_{k=1}^j g_k$. An example is illustrated in Fig.~\ref{fig:1D}, for the case with $n_f=n_g=6$ masses.

The corresponding inverse cumulative distribution functions are also piecewise constant, as can be seen in the top right panel of Fig. \ref{fig:1D}. Note that the figure in the top right panel in Fig. \ref{fig:1D} is the same as that in the top left with the axes swapped.  Specifically we have
\begin{align}
F^{-1}(t) = \begin{cases}x_1&0\le t < F_1\\x_i&F_{i-1}\le t < F_i, \quad (i=2,\dots,n_f)\end{cases}\label{eq:icumul1Df}\\
G^{-1}(t) = \begin{cases}y_1&0\le t < G_1\\y_j&G_{j-1}\le t < G_j, \quad (j=2,\dots,n_g)\end{cases}\label{eq:icumul1Dg} ,
\end{align}
where $t$ is the range variable of the cumulative distribution function (CDF) of $F$ and $G$, $0\le t \le 1$. Evaluation of the expression for the Wasserstein distance in (\ref{eq:solW2}) in this case requires the integral of the difference between two piecewise constant functions, $|F^{-1}(t)-G^{-1}(t)|$, which is itself another piecewise constant function (see the bottom left panel of Fig.~\ref{fig:1D}). 
We write the ordered set of points along the $t$ axis corresponding to the locations of the point masses as $t_k, (k=1,\dots, k_{max})$, ($k_{max}=n_f+n_g-1$).  For ease of notation we assume that the two functions do not share any discontinuities, and $t_{k_{max}}=1$. The $t$-axis of the top right panel in Fig. \ref{fig:1D} shows an example of the relationship between $t_k$ and $(F_i,G_j)$, while the bottom left panel shows the piecewise constant function $|F^{-1}(t)-G^{-1}(t)|$ and its relation to the $t_k$ values for the same case.  For each $t_k$ value we introduce the variable $z$, defined by 
\begin{equation}\label{eq:ijk}
\left.\begin{array}{l}
z^f_k = F^{-1}(t_k)\\
z^g_{k} = G^{-1}(t_k)\end{array}\right\} \quad k=1,\dots,k_{max}
\end{equation}
where the values of $z^f_k$ correspond to one of the point mass locations, $x_i, (i=1,\dots,n_f)$,
and similarly for $z^g_k$ with respect to $y_j, (j=1\dots,n_g)$.  For convenience, we use the functions $i(k)$ and $j(k)$ to describe the mapping of indices and have
\begin{equation}
\left.\begin{array}{l}
z^f_{k} = x_{i(k)}\\
z^g_{k} = y_{j(k)} \end{array}\right\}\quad k=1,\dots,k_{max}\,.
\label{eq:ijkf}
\end{equation}
Inspection of Fig \ref{fig:1D} will help make these relationships clearer.
The $p$-Wasserstein distance for the point masses from (\ref{eq:solW2}) corresponds to the area under the curve shown in the bottom left panel of Fig. \ref{fig:1D}, which can then be conveniently expressed in terms of the values of $t_k$ and $z_k$. We have
\begin{align}
W_p^p(f,g) &=\int_0^1 |F^{-1}-G^{-1}|^p \mathrm{d}t \nonumber\\
&= \sum_{k=1}^{k_{max}} \int_{t_{k-1}}^{t_{k}} |z^f_{k}-z^g_{k}|^p \,\mathrm{d}t\nonumber\\
&= \sum_{k=1}^{k_{max}} |z^f_{k}-z^g_{k}|^p (t_k - t_{k-1}),
\label{eq:solpm}
\end{align}
where $t_0 = 0$. 
We therefore see that for the case where $f$ and $g$ are represented by 1D point masses, evaluation of the $p$-Wasserstein distance reduces to the dot product 
\begin{equation}
W_p^p(f,g) =  \Delta {\bf  z}^T A {\bf t},
\label{eq:Wp}
\end{equation}
where  $ \Delta {\bf  z}^T = (|z^f_1-z^g_1|^p,\dots,|z^f_{k_{max}}-z^g_{k_{max}}|^p)$, depends only on the point mass locations, $(x_i,y_j)$, and  $ {\bf  t} = [t_1,t_2,\dots,t_{k_{max}}]^T$ depends only on the point mass weights, $(f_i,g_j)$; and $A$ is the $k_{max} \times k_{max}$ matrix with unit diagonal and -1 on the sub-diagonal. Note that (\ref{eq:Wp}) is exact for this problem and  involves only one ordering of $k_{max}$ values and a dot product.  The computational cost is therefore $O\left((n^f + n^g) \log (n^f + n^g)\right)$, which is extremely efficient in comparison to a Linear Programming solution that is typically $O\left((n^f + n^g)^3\right)$. We are not aware of this result appearing elsewhere in the literature, and its computational convenience underpins the efficiency of our approach.

\subsection{Calculating Transport Maps} 
In evaluating  the Wasserstein distance for 1D problems,  it has not been necessary to first calculate the Transport Map that relocates $f(x)$ onto $g(y)$. However it can be constructed from the solutions already obtained. 
For the point mass case an analytical solution for the 1D  transport plan may be obtained by comparing (\ref{eq:solpm}) with the Linear Programming equivalent (eq.~\ref{eq:discreteW}) which by inspection shows that the transport matrix is given by
\begin{equation}
\pi_{i(k),j(k)} =  (t_k - t_{k-1}),\quad (k=1,\dots,k_{max}),
\label{eq:discreteTP}
\end{equation}
where the functions $i(k)$ and $j(k)$ indicate the non-zero components of the transport matrix $\pi_{ij}$ defined by (\ref{eq:ijkf}). For all other elements  $\pi_{ij}=0$. 
In this case the transport plan also becomes piecewise constant and its non-zero elements are represented in the bottom right panel of Fig. \ref{fig:1D}. As can be seen, the particles in the $f$ distribution at $x$ give mass to multiple $g$ particles in the corresponding $y$ interval along the vertical sections of the curve. Similarly, the $g$ particles at $y$ receive mass from multiple $f$ particles in the $x$ interval of the horizontal section. This is an example of mass splitting that is common for discrete problems. By combining previous results it can be shown that the Wasserstein distance may be written in terms of the transport plan as  
\begin{equation}
W_p^p = \sum_{i=1}^{n_f} \sum_{j=1}^{n_g} \pi_{ij} | x_i - y_j |^p.
\label{eq:discreteWp}
\end{equation}
We see then that Wasserstein distance in the 1D discrete case may be conveniently evaluated for any $p$ using (\ref{eq:solpm}) or  (\ref{eq:Wp}), while the transport plan is given by (\ref{eq:discreteTP}). In what follows we use these expressions to calculate the marginal Wasserstein distances
$^tW_p(p^{(t)}_{obs}, p^{(t)}_{pre})$ and  $^uW_p(p^{(u)}_{obs}, p^{(u)}_{pre})$.

\subsection{Combining derivatives}\label{sec:chain}
For completeness we briefly discuss the combination of the various derivative terms using the chain rule.  Overall, we require the derivative of the Wasserstein distance between marginals, with respect to model parameters, $m_l, (l=1,\dots,M)$. From (\ref{eq:Wopt}) we have
\begin{equation}
\frac {\partial W_p^p}{\partial m_l} =  \alpha \frac {\partial ^tW_p^p}{\partial m_l} + (1-\alpha)\frac {\partial ^uW_p^p}{\partial m_l} \quad (l=1,\dots,M).
\label{eq:chain0}
\end{equation}
Using the chain rule over all intermediate variables in the four stages we have
\begin{equation}
\frac {\partial ^uW_p^p}{\partial m_l} = \sum_{i}^{n_t} \sum_{j}^{n_u} \sum_{k}^{N} \frac{\partial ^uW_p^p}{\partial p_{j}^{(u)}} ~\frac{\partial p_{j}^{(u)}}{\partial p_{ij}} ~\frac{\partial p_{ij}} {\partial d_{ij}} ~\frac{\partial d_{ij}}{\partial u^{\prime}_k} ~\frac{\partial u^{\prime}_k}{\partial u_k} ~\frac{\partial u_k}{\partial m_l}
, \quad (l=1,\dots,M).
\label{eq:chainu}
\end{equation}
and
\begin{equation}
\frac {\partial ^tW_p^p}{\partial m_l} = \sum_{i}^{n_t} \sum_{j}^{n_u} \sum_{k}^{N} \frac{\partial ^tW_p^p}{\partial p_{i}^{(t)}} ~\frac{\partial p_{i}^{(t)}}{\partial p_{ij}} ~\frac{\partial p_{ij}} {\partial d_{ij}} ~\frac{\partial d_{ij}}{\partial u^{\prime}_k} ~\frac{\partial u^{\prime}_k}{\partial u_k}  ~\frac{\partial u_k}{\partial m_l}
, \quad (l=1,\dots,M).
\label{eq:chaint}
\end{equation}
where expressions have been simplified by using unnormalised fingerprint density amplitudes as intermediate variables. While the multiple nested summations suggest their evaluation may become computational inefficient, in practice the calculation of the derivatives benefits from the fact that each density amplitude, $p_{ij}$ is at most dependent on two waveform amplitudes $u_k$, and hence all other derivatives are zero. In addition, each marginal amplitude, $p_{i}^{(t)}$ and $p_{j}^{(u)}$ is only a function of the density amplitudes in the corresponding $i$th column or $j$th row, respectively. Again all other derivatives are zero. This simplifies the implementation of the chain rule and in practice the Wasserstein derivatives are straightforward to calculate, while also being exact for all terms once derivatives of the forward problem are known.

\subsection{Combining the four stages}\label{sec:algcomb}
Having described all for steps in our calculation of Wasserstein distances, it should now be clear to the reader that each set of output variables at one stage becomes a set of input variables for the next, starting with waveform amplitudes, $(u_{obs},u_{pre})$, and ending with Wasserstein distances for each pair of marginals, $^tW^p_p$ and $^uW^p_p$. While algebraically tedious, this process is nevertheless very convenient, in that it involves only the combination of exactly evaluable expressions. In addition, since we also have analytical expressions for all derivatives they too may be exactly evaluated for each stage and then combined via the chain rule of partial differentiation. Fig. \ref{fig:stages} shows a flow map of the four stages, from right to left, mirroring the order of the chain rule expressions (\ref{eq:chainu}) and (\ref{eq:chaint}).  The far right hand box represents the calculation of the predicted waveforms from model parameters, the derivatives of which are assumed to be available and depend on the forward model.  Each subsequent coloured box represents one of the four stages described above, resulting in Wasserstein distances for each marginal with respect to model parameters. In each box the relevant partial derivatives are also highlighted. 

In summary the parameter changes from right to left read as follows:
%
\begin{enumerate}
\item {\sl Model parameters to waveform}: The solution to the forward problem takes the model parameters, $m_l, (l=1,\dots,M)$ and produces a seismic waveform, $u_i, (k=1,\dots,n)$. Partial derivatives of the waveform amplitudes with respect to the model parameters are also calculated $\partial u_k/\partial m_l$. The details of these calculations depend on the nature of the forward problem and are assumed to be given.

\item {\sl Waveform to non-dimensional space}: In the second box observed and predicted waveforms, $(u_k, t_k), (k=1,\dots,n)$, are transformed to $(u^{\prime}_k, t^{\prime}_k), (k=1,\dots,n)$ using the time-amplitude transformation given by  (\ref{eq:affine})-(\ref{eq:utrans1}).

\item {\sl Non-dimensional waveform to 2D Fingerprint}: In the third box the waveform, $(u^{\prime}_k, t^{\prime}_k), (k=1,\dots,n)$ is used to calculate the nearest distance $d_{ij} (i=1,\dots,n_t; j=1,\dots,n_u)$ to each grid node in the time-amplitude window. 
Calculation details and derivatives appear above and in appendix \ref{app:FP}.

\item {\sl Fingerprint function to Fingerprint density}: In the fourth box the 2D distance field is converted to a density and normalized using (\ref{eq:pdfg}) to (\ref{eq:pnormg}). 
Here the 1D  time and amplitude marginals, $p_i^{(t)}$ and $p_j^{(u)}$ are also calculated from the 2D densities. Details and derivatives appear above and in appendices  \ref{app:Wderiv} and \ref{app:FP}. 

\item  {\sl Fingerprint densities to Wasserstein distance}: In the final box the Wasserstein distances between 1D marginals are calculated. Details of the Wasserstein calculation are given in section \ref{subsec:dpm}, with derivatives in appendix \ref{app:margwass}. 

\end{enumerate}

\begin{figure}
\begin{center}
\includegraphics[width=\textwidth]{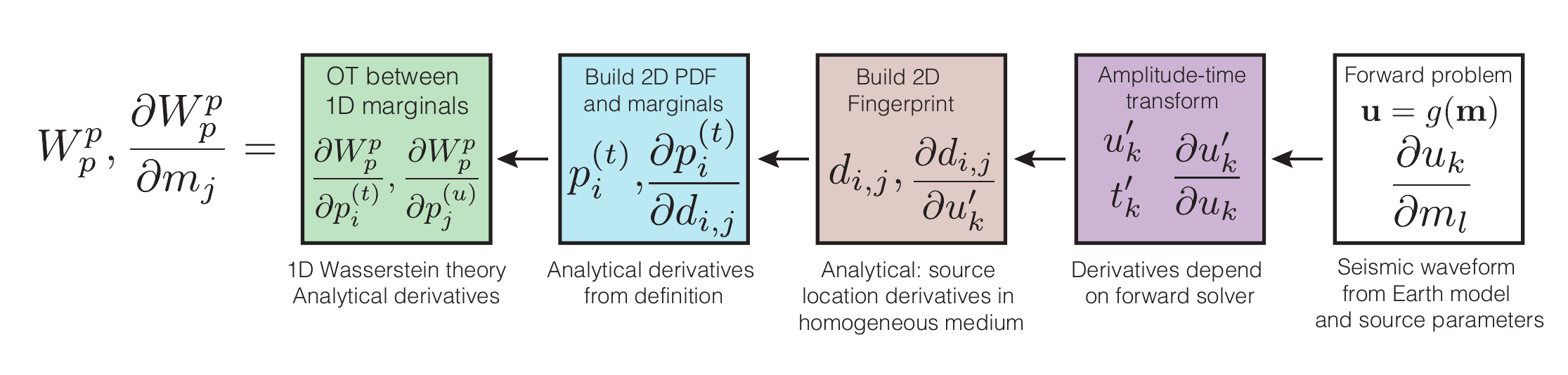}
\end{center}
\caption{\label{fig:stages} A flow map from right to left for the calculation of the Wasserstein distance and its derivatives showing the four stages with their intermediate variables. This ordering mirrors that of the chain rule connecting derivatives in (\ref{eq:chainu}) and (\ref{eq:chaint}). The box on the far right represents the forward problem applied to the model unknowns, $m_l$. The next (purple) box represents a time-amplitude window transformation to non-dimensional variables. In each box a new set of variables is calculated from the old, culminating in the Wasserstein distance between an observed and predicted waveform. Derivatives between intermediate variables are noted in each box. Details of each step in the process appear in the main text. This procedure is used in the examples in section \ref{sec:results}.}
\end{figure}

The reader will observe that nothing in our approach to Optimal Transport of seismic waveforms is dependent on the details of the forward model. 
We simply assume that one exists and derivatives of waveform amplitudes with respect to model parameters are available. Hence the same formulation may be applied to any optimization problem where predicted time series depend on model parameters and must be fit to observations. Such problems occur across the physical sciences and so the formulation here may have general applicability. Furthermore, we note that derivatives of waveform amplitudes with respect to model parameters may not be available in all cases. Indeed it is common in many large-scale problems to use adjoint methods to directly evaluate derivatives of a waveform misfit function with respect to Earth model parameters, e.g. in global waveform inversion \citep{Tromp:2005,Sieminski:2006}. Adjoint methods have previously been used for derivative calculations in Optimal Transport, most notably in exploration seismology \citep{Engquist:2014, Engquist_etal:2016}, although, as noted at the outset, those formulations different from the one proposed here and so it can not be assumed that methods simply carry over.  Formulation of an adjoint approach for our marginal Wasserstein distances is a straightforward extension.

\begin{figure}
\begin{center}
\includegraphics[width=0.5\textwidth]{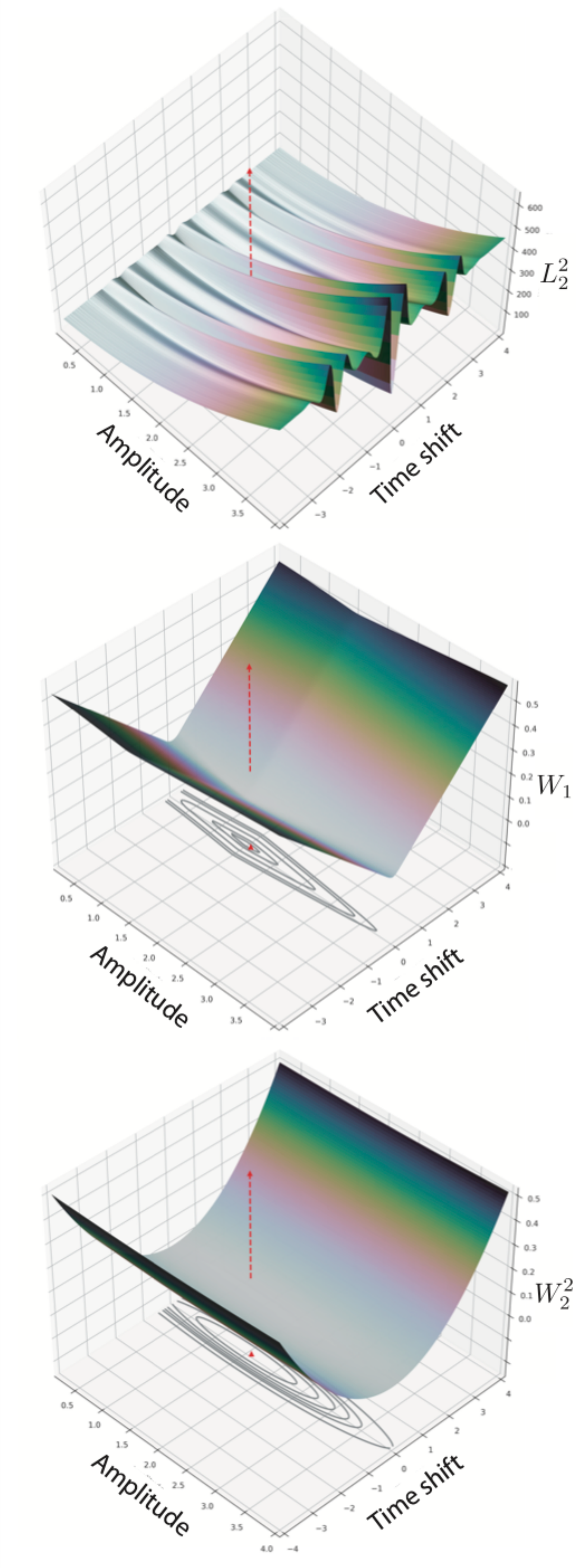}
\end{center}
\caption{\label{fig:rickermisfit} Misfit surfaces between double Ricker wavelets as a function of amplitude and time shift parameters (see text). Top panel shows the multimodal least squares case. The middle and lower panels show the Wasserstein $W_1$ and $W_2^2$ misfits respectively. These were calculated using the marginal algorithm described in the text, and are globally piecewise linear and quadratic respectively. The red dashed line is the global minimum and misfit contours are plotted on the base plane.}
\end{figure}

\begin{figure}
\begin{center}
\includegraphics[width=0.9\textwidth]{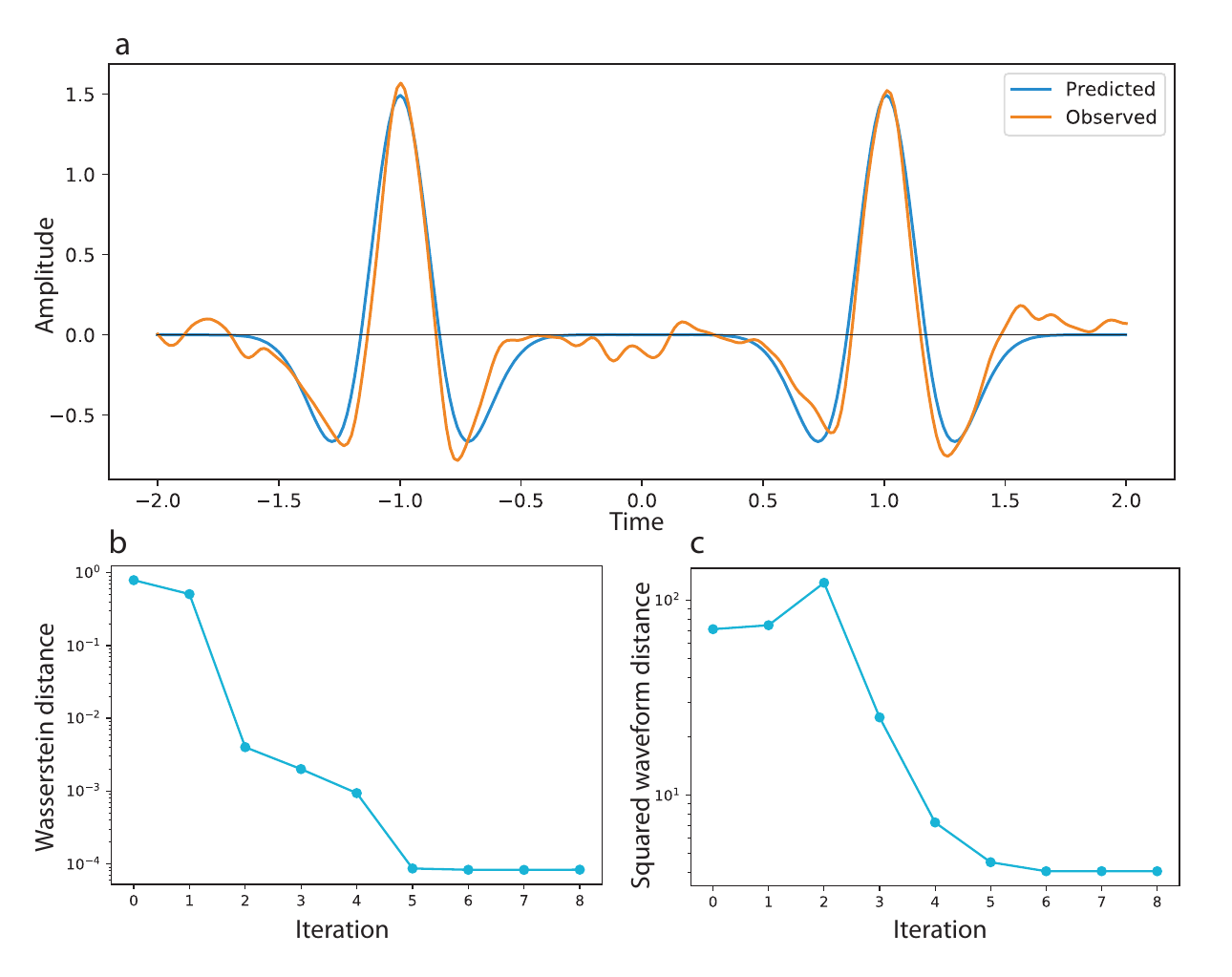}
\end{center}
\caption{\label{fig:rickerfresults} a) shows the best fit of the noisy Ricker wavelets found from minimization of the Wasserstein distance, $W_2^2$. The initial position of the predicted wavelet is shown in Fig. \ref{fig:rickerfp}b, and has had its time, amplitude and frequency parameters optimally adjusted to fit the target waveform, Fig. \ref{fig:rickerfp}a, by gradient minimization. Recall that 5\% Gaussian noise is added to the orange curve. Panels b) and c) show the Wasserstein and the least squares misfits, respectively, at each iteration of  the gradient-based  minimization of  $W_2^2$. As the Wasserstein distance is progressively minimized the $L_2^2$ misfit values, of the same models, show a jump over a local maximum, demonstrating the benefit of the Wasserstein distance function that has only a global minimum for this problem (as seen in Fig. \ref{fig:rickermisfit}). Note the logarithmic scale in (b) and (c).}
\end{figure}

\vspace{10pt}
\section{Examples}\label{sec:results}

\subsection{A toy problem: Double Ricker wavelet fitting}\label{sec:drickerfit}

To demonstrate the utility of this marginal Wasserstein waveform misfit function, we first consider a simple toy problem that is a modification of one studied by \cite{Engquist:2014}, consisting of the fitting of a double Ricker wavelet with added correlated Gaussian noise. We create a three parameter waveform fitting problem by writing a double-Ricker function as
\begin{equation}
r(t) = A[1-2\pi^2f_0^2(t-t_1)^2] e^{-\pi^2 f_0^2 (t-t_1)^2} + A[1-2\pi^2f_0^2(t-t_2)^2] e^{-\pi^2 f_0^2 (t-t_2)^2}
\label{eq:dricker}
\end{equation}
where $t_1 = (t_0 - L/2)$, $t_2 = (t_0 + L/2)$, and $2L$ is the total time window. The three  controlling parameters are the amplitude scale factor, $A$, the time shift, $t_0$, and the reference frequency,  $f_0$. Together these specify the shape and position of the wavelet. The top panel of Fig.~\ref{fig:ricker_align} (orange curve) and Fig. \ref{fig:rickerfp}a show examples with Gaussian random noise added. For the latter case the corresponding parameter values are $L=2$, $t_0 = 0, A = 1.6, f_0=1.0$, which serves as a reference `observed' waveform in our synthetic tests, that is to be fit by locating the optimal values of the triplet $(A,t_0,f_0)$. An example of a sub-optimal fit is the blue curve in the second panel of Fig.  \ref{fig:ricker_align}, which has the correct amplitude, $A$ and frequency, $f_0$ but incorrect time shift, $t_0$. Another is seen in Fig.~\ref{fig:rickerfp}b which has too low an $A$ and $f_0$, but too large $t_0$.  

We perform three separate experiments, the results of which are summarized in Figs. \ref{fig:ricker_align}, \ref{fig:rickermisfit} and \ref{fig:rickerfresults}, respectively. The first mirrors the problem studied by \citet{Engquist:2014}, but uses our marginal Wasserstein distance algorithm as described in section \ref{sec:alg}.  We sweep through a single axis in our three parameter model space, by varying the time shift parameter, $t_0$, with the amplitude and frequency parameters fixed at their true values. The third panel in Fig. \ref{fig:ricker_align} shows the usual summed squared difference in waveform amplitudes, $L_2^2$, as a function of $t_0$. As one might expect this misfit function has multiple local minima (at least 11 in this case), caused by incorrectly aligned peaks in the double-Ricker wavelets. The bottom panel in  Fig.~\ref{fig:ricker_align} shows the marginal Wasserstein distances, $W_1$ and $W_2^2$ calculated with our algorithm, using a value of $\alpha = 0.5$ in (\ref{eq:Wopt}).
The parameters used are, $\Delta_t = 4$s, with $\Delta_u = 1.2\Delta_a$, a 20\% increase of the amplitude range as above; $n_u=80, n_t=512$ for the fingerprint grid, and $s=0.03$ for the distance scale factor.
As can be seen both the $W_1$ and $W_2^2$ misfit functions have a single global minimum at the true solution, with the former having a discontinuous gradient at the optimum. They are reminiscent of a simple $L_1$ and $L_2$ norm measure in model space, which is very encouraging. A figure very similar to the $W_2^2$ curve was obtained by  \citet{Engquist:2014} for their Wasserstein distance tests and we can conclude here that our algorithm reproduces the same features in this case, even in the presence of noise on the signal.

As a second experiment we extend this model space sweep by comparing misfit surfaces for $L_2^2$, $W_1$ and $W_2^2$ as a function of amplitude and time shift parameters. These are displayed in Fig.~\ref{fig:rickermisfit}. The $L_2^2$ surface exhibits ridges of local maxima and troughs of local minima extending the local minima seen in Fig.~\ref{fig:ricker_align} along the amplitude axis. The Wasserstein misfit surfaces for $W_1$ and $W_2^2$ show global uniformity for this problem, with the former again being similar to a series of linear segmented planes, and the latter apparently a global quadratic-like surface. As one would expect, the global minimum is not exactly at the true solution, given by the red dashed line, due to the presence of Gaussian noise added to the `observed' Ricker function.

As a third experiment, we performed a derivative based optimization of both the $L_2^2$ and the Wasserstein measure, $W_2^2$, for all three model parameters $(a, t_0, f_0)$. 
For the $L_2^2$ case we calculated exact derivatives by differentiating (\ref{eq:dricker}) and using the usual definition of a squared waveform difference. For the Wasserstein case we used the same Ricker wavelet derivatives and combined them with (\ref{eq:chainu}) and (\ref{eq:chaint}) to get derivatives of the summed marginal Wasserstein distances (\ref{eq:chain0}) with respect to the three unknowns. Fig.~\ref{fig:rickerfresults}a shows the best fit curves after 8 iterations minimizing $W_2^2$ with a Limited-Memory  Bound-Constrained  gradient descent algorithm (L-BFGS-B)\citep{Byrd:1995,Zhu:1997,Virtanen:2020}.  The starting Ricker waveform for this minimization is shown in Fig. \ref{fig:rickerfp}b.  Fig.~\ref{fig:rickerfresults}b shows the reduction on $W_2^2$ at each iteration. Clearly, the solution is a close fit and convergence to the global minimum has been achieved in all three parameters. The same experiment was performed with the $L_2^2$ misfit, which converged to a secondary minimum. To allow some comparison between $W_2^2$ and $L_2^2$ we repeated the $W_2^2$ optimization and at each iteration calculated both the $W_2^2$ and $L_2^2$ misfits of waveforms. These curves are shown in Fig.~\ref{fig:rickerfresults}b--c. Notice that the $L_2^2$ curve initially increases before decreasing, indicating that it has been necessary to traverse over one of the ridges in Fig.~\ref{fig:rickermisfit} to find the global minimum, which  explains why the $L_2^2$ optimization failed; the $W_2^2$ optimization does not encounter this ridge and converges smoothly.

The three Ricker waveform fitting experiments provide an encouraging set of results, albeit in a simple problem, suggesting the potential of Wasserstein distance for waveform fitting generally. The following experiment also provides a demonstration of the utility of the derivative calculations described in section \ref{sec:alg}.

\subsection{Coupled source and moment tensor inversion of displacement seismograms}\label{sec:moment}
For a more comprehensive---and geophysical---test we applied our algorithm to the problem of coupled moment tensor and source location inversion. Here the aim is to compare  performance of both $L_2^2$ and $W_2^2$ in a more complex seismic waveform fitting problem, where exact solutions are known. To this end we replicate the problem of fitting high rate GPS displacement waveforms studied by \citet{OToole2012} for the moment tensor and  spatial location of the $M_w$ 6.6 2005 Fukuoka earthquake. 

\begin{figure}
\begin{center}
\includegraphics[width=0.5\textwidth]{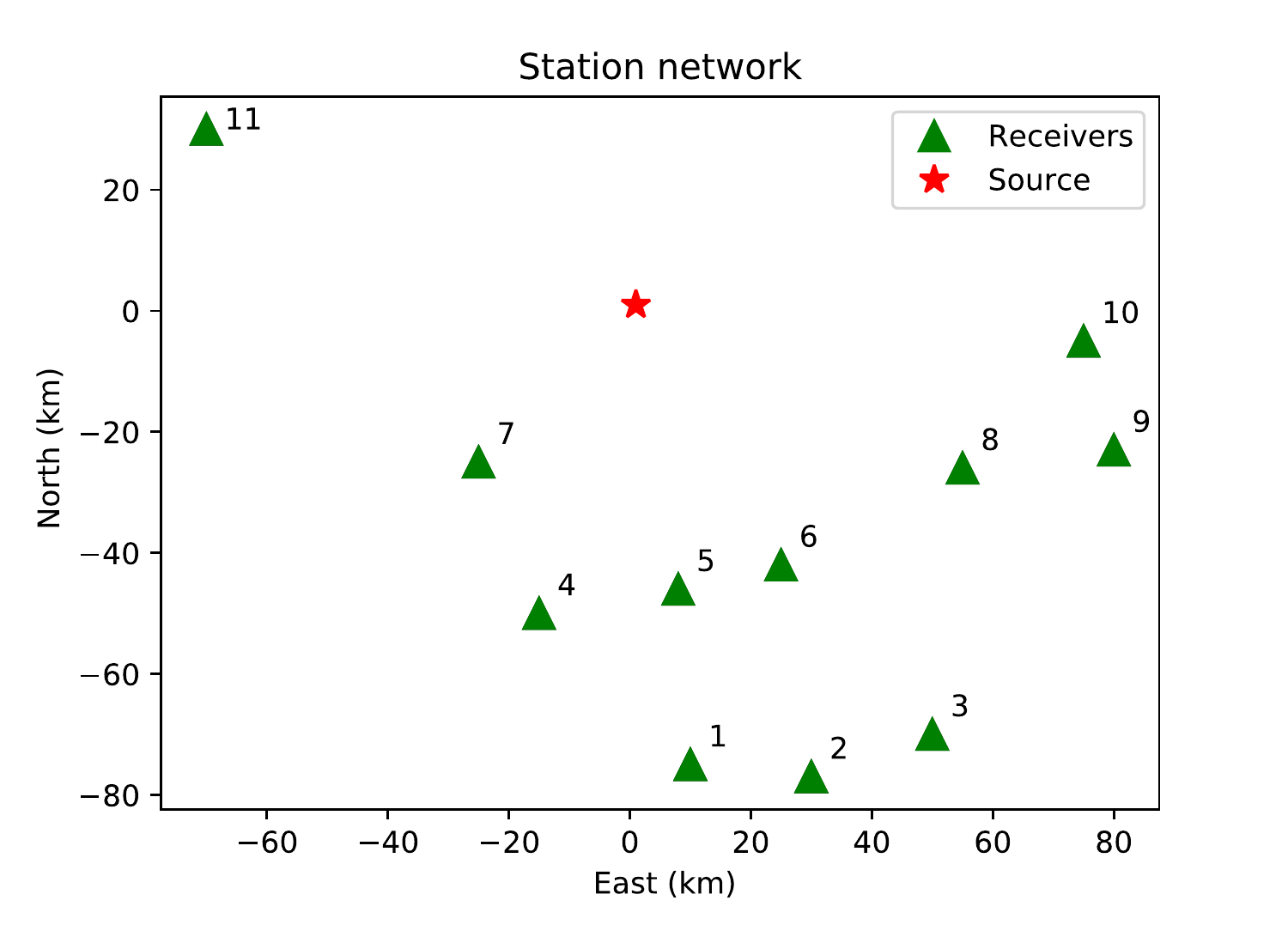}
\end{center}
\caption{\label{fig:stations} Station network (triangles) used in the source inversion experiments. These correspond to a Cartesian projection of the actual GEONET GPS receivers  used by \citet{OToole2012} to invert displacement waveforms of 2005 Fukuoka earthquake (red star). }
\end{figure}

Fig. \ref{fig:stations} shows the location of 11 GEONET stations used in our experiment, relative to a source (shown as a star at $x=1$, $y=1$, $z=20$) in a local Cartesian framework. For each station we calculated the three component displacement seismograms shown in Fig. \ref{fig:seismograms}a with 1s sampling rate for 60s duration and origin time at $t=0$s. Waveforms were calculated with an implementation of the algorithm of \citet{OToole2011} which produces numerically stable computation of complete synthetic seismograms including the static displacement in plane layered media \citep{Valentine:2021}. All seismograms were calculated using the moment tensor solution obtained by \citet{OToole2012} and in the six layered seismic velocity model used by  \citet{Kobayashi:2006} to study this earthquake \citep[see Table 1 of ][]{OToole2012}.
As shown in Fig. \ref{fig:seismograms}a, the solid red, blue and green component  traces have Gaussian correlated random noise added, with amplitude equal to 6\% of the maximum seismogram amplitude in each window, and a temporal correlation length of 5s (i.e. 1/12 of the window length). Note the static offset present in many cases. These are taken as `observed' seismograms for our synthetic experiments.

In all experiments an $L_2^2$ waveform misfit was calculated by summing the squared amplitude difference between all 33 pairs of observed and predicted waveforms. Time-amplitude windows use the parameters $\Delta_t = 60s, \Delta_u = \pm 30$\% of the  amplitude range of the observed waveform in each window. We calculated the two marginal Wasserstein distances, $^tW_2^2$, i.e. with $\alpha=1$ in (\ref{eq:Wopt}) and $^uW_2^2$  with $\alpha=0$ in (\ref{eq:Wopt}), with the fingerprint grid, $n_t = 61,n_u = 79$ and distance scale factor $s=0.04$.  The method of \citet{OToole2012} was used to calculate exact derivatives of displacement seismogram amplitudes with respect to source location, structural parameters or moment tensor components. Together with the theory of section \ref{sec:alg} this means that derivatives of both the $L_2^2$ waveform misfit and $W_2^2$ may be calculated exactly for all source locations and moment tensors.  

\begin{figure}
\begin{center}
\includegraphics[width=0.7\textwidth]{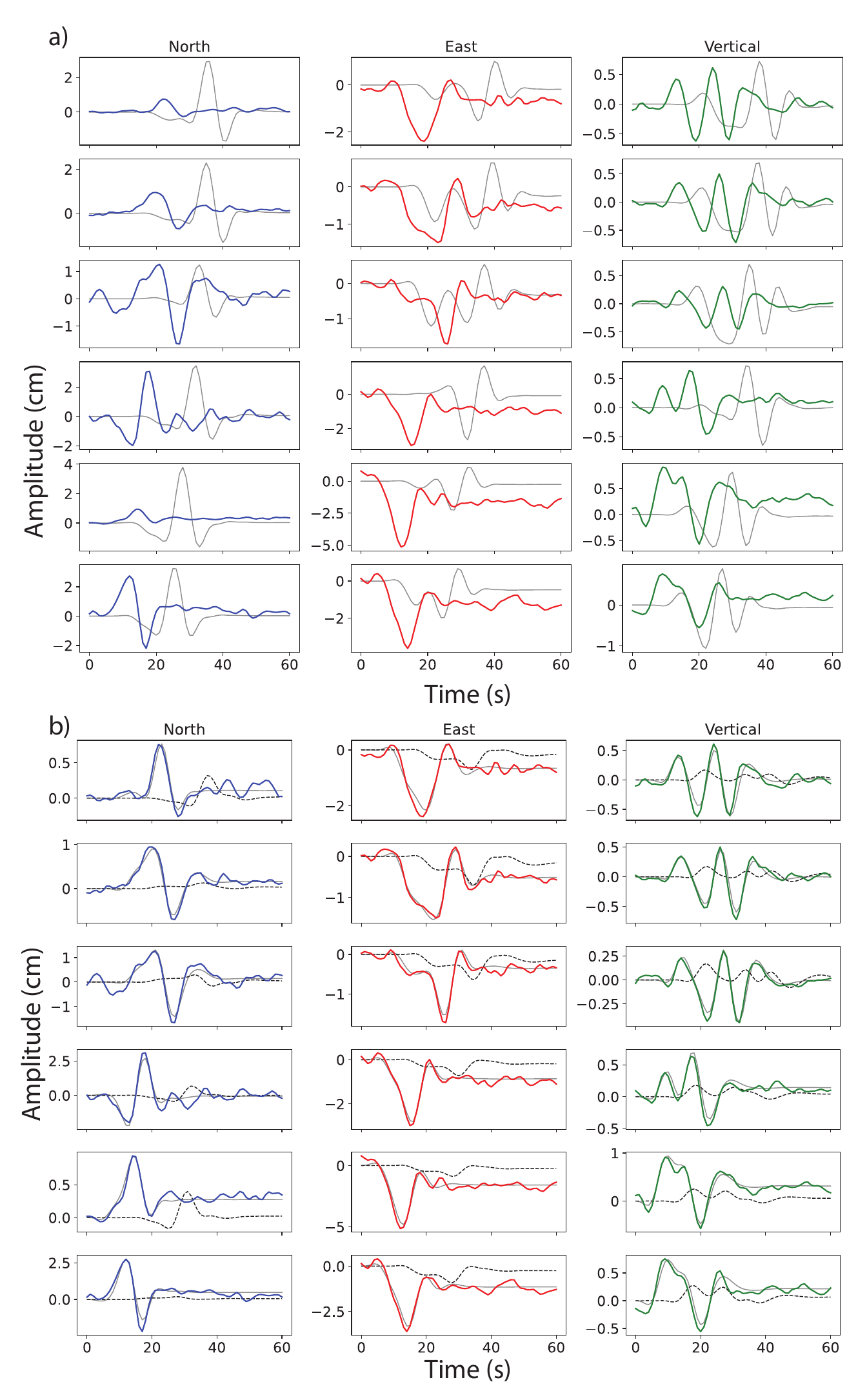}
\end{center}
\caption{\label{fig:seismograms} 
a) shows `observed' three component noisy displacement waveforms (solid, coloured) from a source at location $(1.0,1.0,20.0)$ and predicted displacement waveforms (solid, gray) for an initial guess location 56 km away at $(40.0,40.0,10.0)$. Six of the eleven receivers are shown, numbered 1-6 in Fig. \ref{fig:stations}. b) shows `observed' and best fit displacement waveforms for the same receivers obtained from the least squares misfit (dashed, black) and Wasserstein misfit (solid, gray).
For the Wasserstein misfit the minimization converges to the global minimum, which is within 2.1 km from the true location and all waveforms are fit well, whereas the least squares case converges to a location $\approx$ 67 km from the true location and discrepancies can be seen for many receiver components pairs.}
\end{figure}

We performed three experiments with this realisation. The first was to calculate 2D planar cross sections of $L_2^2$ and Wasserstein misfit functions for different depths and inspect them for local minima. Fig. \ref{fig:misfits} shows four depth slices for both $L_2^2$ and $^tW_2^2$ across a $\pm 60 $ km region about the true source location (marked with a black dot). These experiments were performed for a source depth at $z= 20 $ km, with the origin time and moment tensor set to the true value. One can see that the true source location does not lie at the global minimum of either misfit function due to the random noise added to the observed waveforms. The $L_2^2$ misfit contours are characterised by being predominantly a single global minimum below the true source depth of $20 $ km, although there is a broad secondary minimum in the North of the region which is more apparent at shallower depths. Multiple local minima are clearly observable in and above the $20 $ km layer, especially at the shallowest depth, where more than a dozen minima are apparent. This suggests that gradient based optimization of the $L_2^2$ misfit function may be prone to entrapment in local minima, depending on starting location of an iterative sequence. By contrast, the Wasserstein misfit function is similar in all four depth layers, with a clear dominant global minimum in each case and no significant local minima. 

In an attempt to quantify the  reduction in the number of local minima observed in the Wasserstein surface, we calculated  the values of $L_2^2$ and $^tW_2^2$ over a 100x100 grid across the two shallowest layers shown in Fig. \ref{fig:misfits}, and identified all points for which the misfit increased in each of the North, South, East and West directions. For $L_2^2$, 37 and 20 points met this criterion in the first two layers respectively, i.e. Figs. \ref{fig:misfits}a and \ref{fig:misfits}b, while for $^tW_2^2$, 5 points had this property in each of the corresponding surfaces,  i.e. Figs. \ref{fig:misfits}e and \ref{fig:misfits}f.   This is encouraging, and suggests that optimization of the Wasserstein based misfit function may be more successful than of the standard $L_2^2$ norm in terms of avoiding entrapment in local minima. This is also consistent with the observations of \citet{Hedjazian:2019} in their comparison of Wasserstein misfits with $L_2^2$ for seismic receiver functions. 

In the next experiment we performed multiple gradient based inversions with both $L_2^2$ and $W_2^2$, with $\alpha=1/2$, from different initial source locations, while retaining the correct moment tensor parameters.  Fig. \ref{fig:seismograms}a shows the initial displacement waveforms, in gray, for the starting source. One can see that the initial phases of the predicted waveforms are not aligned with those of the observed, indicating that the starting locations are relatively distant from the true source, and that gradient based methods may suffer from `cycle skipping'.  In addition, since these are displacement rather than velocity seismograms and include static offset, we can see that most of the predicted waveforms from this source also have differences in their static offset. The optimisation therefore needs to fit amplitudes by both aligning phases and correcting static offsets. 

\setcounter{table}{0}
\begin{table*}
\begin{center}

\label{tab1}
\begin{tabular}{@{}llllcll}
  Location & x & y & z & $\Delta D$\\ 
\hline
True & 1 & 1 & 20 \\
Start & 40 & 40 & 10 & 56.1\\
$W_2^2$ & 0.6 & 1.9& 21.9 & 2.1\\
$L^2_2$ & 5.5 & 58.2 & 54.8 & 67.1 \\
\end{tabular}
\caption{Source location results for test problem in Fig. \ref{fig:seismograms}. $\Delta D$ is the distance to the true source location. All values in kms.}
\end{center}
\label{table:locations}
\end{table*}

Fig. \ref{fig:seismograms}b  shows the same observed seismograms together with the predictions from the best fit solution produced by optimising $L_2^2$ (black dashed lines) and $W_2^2$ (coloured dashed lines), respectively. While the former has aligned the waveforms somewhat better than the initial set, the fits remain quite poor in many cases, both in the phase alignment and the static offset, whereas minimization of the Wasserstein misfit has resulted in good alignment of all phases and static offsets. In this case the Wasserstein minimization converged to a location $2.1$ km from the true source, consistent with being at the global minimum of the misfit function in Fig. \ref{fig:misfits}f. Minimization of $L_2^2$ on the other hand converged to a location $67$ km from the true source suggesting entrapment in a local minima. Full results are given in Table 1.

This test was repeated using a suite of 48 different starting locations, spread regularly across the 3D volume comprising $\pm 60 $ km laterally and $0-40 $ km in depth. At each of the four depths $10,20,30,$ and $40  $ km, 6 starting locations were arranged along both NW-SE and NE-SW diagonals at co-ordinates of $\pm 60, 40, 20$ km respectively, and these are shown as white circles in Fig. \ref{fig:misfits}d. 
Results of the suite of repeat minimizations of both $L_2^2$ and $W_2^2$ from the 48 initial locations are summarized in Fig. \ref{fig:sourcecmt}a. Here we plot the solution error for minimization of $W_2^2$ (x-axis) versus $L_2^2$ (y-axis). We observe that minimization of the Wasserstein misfit results in many more cases of convergence to an effective global minimum than the L$_2^2$ case from the same starting location. Here we define convergence as meaning the final location is within $2.5$ km of the true location, which is indicated as dashed lines in Fig. \ref{fig:sourcecmt}a.  We found that 77\% of the trials converged to the global minimum using the Wasserstein misfit and 41\% with the $L_2^2$ misfit. There are also cases where both converged, or neither converged, but none where the $L_2^2$ converged and the $W_2^2$ did not.  In assessing these results, it must be remembered that the failure of minimization algorithms to converge to a global minimum may be due to several factors including the accuracy of the derivatives provided and the details of the optimization procedure itself. We are confident that the first count is not significant here as independent experiments show that accurate derivatives are calculated in both cases. On the second count, we merely rely on the fact that we use the same L-BFGS-B algorithm in each case.

In the final experiment we extend the previous result to the case where moment tensor components are also included. These six moment tensor parameters are linearly related to displacement waveform amplitudes and so this addition increases the size of the model space and creates a mixed linear-nonlinear problem. We repeat the misfit minimizations for both $L_2^2$ and $W_2^2$ from the 48 source initial locations with initial moment tensor parameters determined by linear inversion. Some preconditioning of the gradient was required in both cases to balance the relative influence of source location and moment tensor components. Fig. \ref{fig:sourcecmt}b shows the results in a similar format to the location-only optimization. 
Here the global minimum is about $0.6$km from the true solution in the $(x,y,z)$ co-ordinates, and we define convergence as meaning the final location is within $1.0$ km of the true location, which is indicated as dashed lines in Fig. \ref{fig:sourcecmt}b.
As with the earlier experiment, the results show the clear improvement in convergence when using $W_2^2$ over the $L_2^2$. For this case 54\% of the trials converged to the global minimum using the Wasserstein misfit and 35\% with the $L_2^2$ misfit. This case shows the least improvement of Wasserstein over $L_2$, and may reflect the poorer performance of the optimisation method in the larger dimensional space. In addition it may be worthwhile exploring the influence of the various scaling parameters, $(\alpha,\Delta_u, s)$, on the character of Wasserstein misfit, which we leave for future study. 
\begin{figure}
\begin{center}
\includegraphics[width=0.8\textwidth]{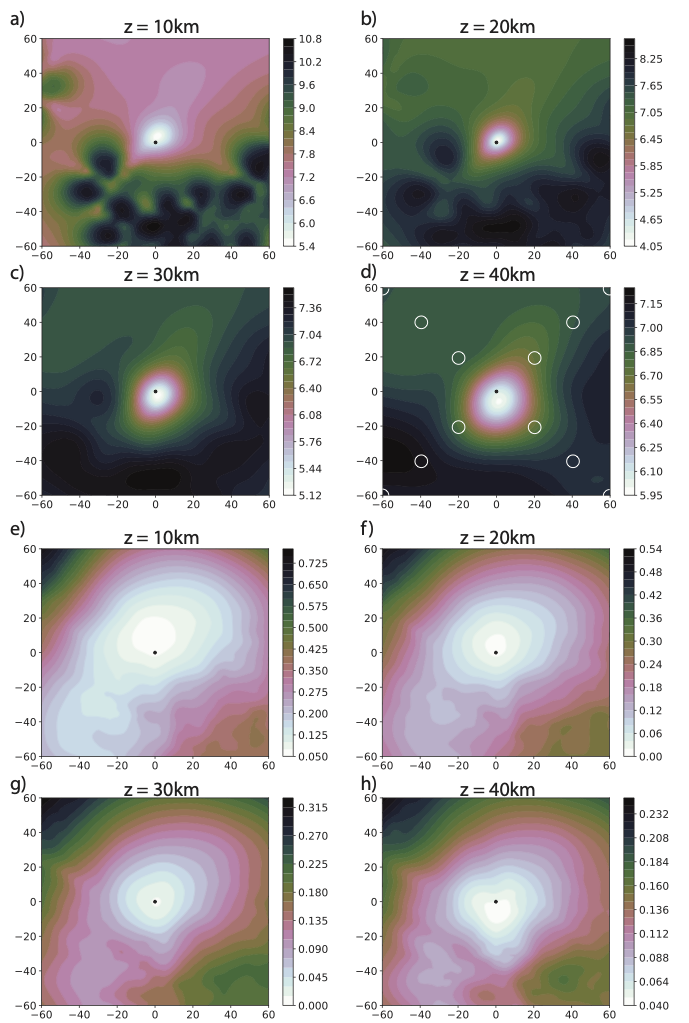}
\end{center}
\caption{\label{fig:misfits} Contoured slices through Least squares, $L_2^2$, (panels a-d) and Wasserstein misfits, $W_2^2$, (panels e-h) for the source relocation experiments. For each misfit four depth slices are shown (10-40 km) over a $\pm 60$  km lateral range about the true source, given by the black circle. White circles in panel d show the lateral location of initial sources used in the repeat optimization test.  Colour scales span misfit values in each slice, given by the legend. The true source is at 20 km depth. Overall the Least squares has few secondary minima below the source depth, but has 11 or more local minima at or above the source depth. The Wasserstein exhibits less variation between depths, a more quadratic appearance and fewer local minima.}
\end{figure} 
\begin{figure}
\begin{center}
\includegraphics[width=0.9\textwidth]{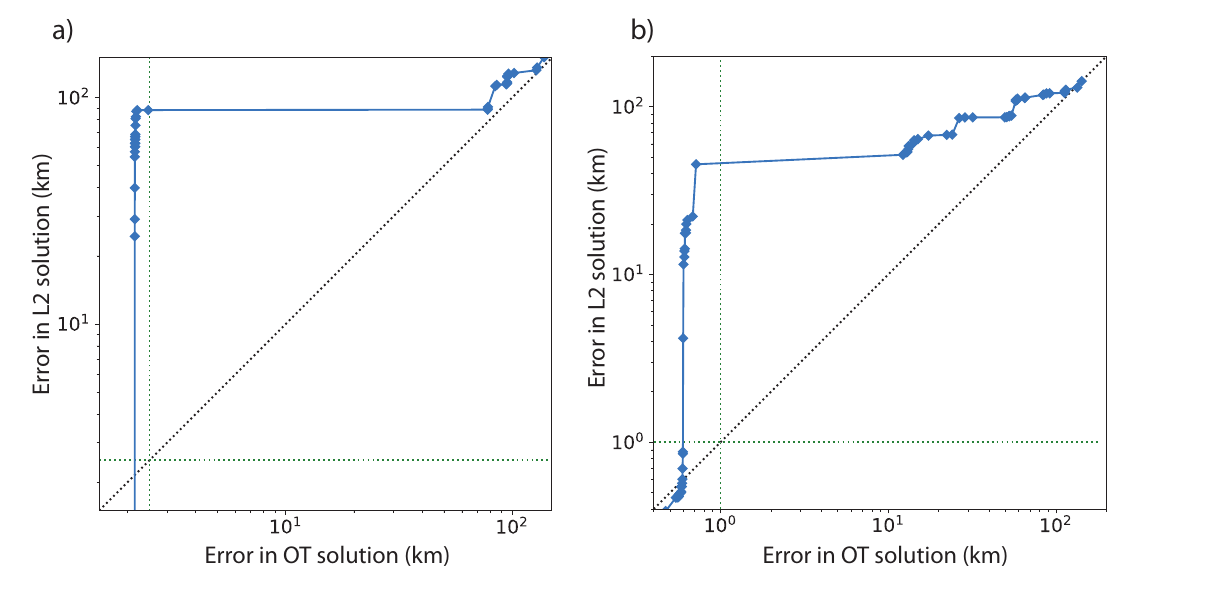}
\end{center}
\caption{\label{fig:sourcecmt} Errors in source location obtained by minimizing a least squares waveform misfit, $L_2^2$,  compared to a Wasserstein distance, $W_2^2$, for 48 cases with different starting locations up to 86  km laterally and 20 km in depth from the true source location. a) is for source location keeping moment tensor fixed at the true value, while b) is for coupled source location and moment tensor inversion. Dashed lines indicate a 2.5 km (a) and 1 km (b) error from true location, respectively, which are used as criteria for convergence in the main text. In both cases the Wasserstein results in convergence to the global minimum in a greater number of cases than the least squares misfit.}
\end{figure}

An area not yet addressed is the computational efficiency of our approach.  In our experiments the calculation of the 2D density function (the `fingerprint') for a single displacement waveform in Fig. \ref{fig:seismograms} took 0.016s of CPU. This used a grid of $n_t=61, n_u= 79$ and was performed with a Python 3 script on a Macbook Pro with 2.9Ghz Quad-Core Intel Core i7 processor and 16Gb memory. In the same experiment, the cost of an evaluation of all 33 displacement waveforms including derivatives and the $L_2^2$ misfit using our (Python) implementation of the waveform forward solver of \citet{OToole2012}, was 2.37s.  This compares to the same calculation for the $W_2^2$ Wasserstein misfit of 2.89s. While these values depend on implementation details, their ratio is perhaps more robust, and so we observe just a 22\% increase for the Wasserstein calculation over that of $L_2^2$ including derivatives. Given that this involves all 5 stages in Fig. \ref{fig:stages}, it does not seem a significant burden. These relative computation costs carry over to a complete execution of the optimization algorithm. For example, in the source location experiment shown in Fig. \ref{fig:seismograms}b the minimization of the $L_2^2$ misfit took 37.6s to perform 6 iterations of the L-BFGS-B algorithm, which required 13 misfit and Jacobian evaluations, whereas the equivalent Wasserstein minimization took 51.2 s to perform 9 iterations with 17 misfit and Jacobian evaluations, an increase of 36\%. We also compared the compute cost of our Wasserstein misfit, utilizing marginals, with a fully 2D Wasserstein calculation using the entropically smoothed Sinkhorn algorithm \citep{Cuturi:2013,Flamary:2021}. Here our marginal Wasserstein algorithm took 5666s to compute the 3D misfit grid shown in Fig. \ref{fig:misfits} while the same calculation using the Sinkhorn algorithm was estimated at 11.8 days based on the cpu time for a single Wasserstein distance averaged over 20 trials. Hence there is more than two orders of magnitude in their relative compute costs.
In our experiments, we can therefore conclude that in terms of computational costs, optimization of the marginal Wasserstein is not a significant increase over that of $L_2^2$ and that our marginal Wasserstein algorithm is extremely efficient compared to an exact 2D OT calculation.

Overall we see a consistent picture that the Wasserstein misfit performs better, in that it permits a standard optimization algorithm to more reliably converge to the correct solution without significant additional computational cost. In our experiments it would be feasible, if cumbersome, to have used either misfit together with an enumerative search to find the global minimum. For both misfit functions the global minimum is consistently near the true solution, as one would expect.  Our experiments suggest that when enumerative search is infeasible, due to costs of forward modelling as the number of unknowns increases, then gradient based optimization is apparently more reliable with the Wasserstein misfit. We take this as an encouraging result, and as an indication that gradient-based methods for the nonlinear fitting of other time series might show similar results.

\section{Discussion and Conclusions}\label{sec:discus}

In this paper we have proposed a new approach to the use of Optimal Transport based waveform misfit functions for inversion, and developed computational algorithms to enable this. Our method allows seismic waveforms to be compared in independent, non-overlapping time windows with arbitrary amplitudes. This general setting can be used for regional or global waveform fitting problems, and in principle for any case where two time series need to be compared. The central idea is to avoid the common practice of directly transforming an oscillatory waveform to a positive one, and instead build a 2D density function about each waveform and then calculate Wasserstein distances between their marginals.  We have shown how this approach leads to an entirely analytical formulation, allowing use of 1D Optimal Transport theory and exact calculation of derivatives of the Wasserstein distances with respect to waveform amplitudes.

A first experiment was performed on a toy problem problem fitting two double-Ricker wavelets as a function of time shift, amplitude and  frequency which verify that both the proposed $W_1$ and $W_2^2$ Wasserstein misfits have global linear and quadratic shapes respectively, in this simple case. A second experiment was performed involving the fitting of high rate GPS displacement waveforms for seismic source and moment tensor parameters. Here  the Wasserstein misfit $W_2^2$ was more complicated than in the earlier experiment and the presence of local minima could not be ruled out.  Nevertheless it was clear that the Wasserstein misfit surface again showed less complex structure, without the many local minima observable in the corresponding $L_2^2$ misfit surface.   This conclusion was supported by a significantly improved convergence rate of gradient based optimization algorithms and other analyses. Computational efficiency of the new approach has been investigated and results are also encouraging. They suggest that Optimal Transport based misfit measures for time series need not result in significantly increased computational demands. Together, we feel these experiments help illustrate the power of Optimal Transport based misfit functions and build confidence for their implementation in regional and global seismological applications.

Our approach has been to compare the proposed marginal Wasserstein misfit algorithm for waveforms to a standard $L_2^2$ misfit for synthetic cases where we know the true solution. As noted at the outset, there is a long history of authors proposing alternate misfit measures for exploration and regional seismic waveform inversion. Many of these are aimed at reducing the influence of non linearities and over-coming cycle skipping. We have not  attempted a comparison of our proposed algorithm with the plethora of alternatives in the literature, and so make no comment on how it would compare. We note, however, that recently a review of misfit functions for full seismic waveform inversion has appeared \citep{Pladys:2021}. These authors also perform detailed comparison of the performance of two formulations of Optimal Transport, namely the minimization of the KR-norm \citep{Metivier:2016a} and Graph Space OT \citep{Metivier:2018,Metivier:2019}, with other prominent waveform misfit functions, including those based on instantaneous phase
\citep{Bracewell:1965, Taner:1979,Bozdag:2011,Wu:2014}, the Gabor Transform \citep{Fichtner:2008}, the Normalised Integration Method \citep{Donno:2013}, and Adaptive Waveform Inversion \citep{Warner:2016}, all of which have shown efficacy in overcoming cycle skipping problems.

Here Wasserstein distances are used as an alternative measure for the data misfit part of an objective function only. 
If an inversion setting also required addition of one or more regularization terms to an objective function, then these can be minimized together in the usual way, as data misfit and regularization terms are typically independent calculations. However it is conceivable in some cases that the theory of Optimal Transport might also have application to the calculation of regularization terms as well, which are themselves usually a distance measure in model space, often based on an $L_2$ norm. A recent study by \cite{Engquist:2020}, suggests that use of the $W_2^2$ as a data misfit measure itself has a smoothing effect on the inversion process.

We conclude with some remarks on the relationships between the Optimal Transport approach set out here, and other approaches to geophysical inversion. First, we emphasise that there is no universal solution to an inverse problem: the choice of misfit function defines the sense in which we set out to `explain' observational data, and the `best-fitting' model by one measure need not be optimal by another. In the examples presented here, it happens that the $L^2_2$ and $W^2_2$ global minima are similar, but this is not necessarily universally true. Moreover, readers will note that our construction of the Wasserstein misfit does not allow for any statistical treatment of noise, such as covariance information. This may be an avenue for future development, but it implies that in our current formulation the Wasserstein misfit cannot be used to define a true Likelihood.

The concept of a transport map as a means of connecting two distributions has a range of potential applications. In particular, it may offer a route towards simplifying computations in probabilistic settings: operations may be performed on one distribution and then transported to another. For example, if a transport map were to be constructed between the prior and posterior distributions for a given problem, it would then be possible to generate samples from the (often complex) posterior by sampling the (simpler) prior and then mapping these samples appropriately. This could avoid the need for expensive Monte Carlo sampling, and suggests possible connections to the `prior sampling' framework described by \citet{Kaufl2016a}.

This also points towards an alternative strategy for posing and solving Bayesian inference problems: one could write a posterior in terms of the prior and a parameterised representation of a transport map. Then, one could formulate and solve a variational inference problem to find parameters such that Bayes' Theorem is satisfied. This possibility has been explored by \citet{ElMoselhy:2012}, but does not seem to have received widespread attention.

These are interesting avenues that we believe are worthy of further investigation, and that may provide theoretical and conceptual connections to support the development of the Optimal Transport approach developed in this paper. It is clear that this theory is a powerful approach for quantifying the similarity of density functions, and it can be applied to any inversion or optimisation problem that has this form. In this paper, we have focussed particularly on time series data, and proposed a strategy for converting these into a form amenable to application of Optimal Transport approaches. This enables waveform similarity to be assessed in terms of a measure of the `work' required to transform one signal into the other, and we have shown that this can perform well in optimisation settings compared to conventional $L_2$ misfits. While we bear in mind the famous `No Free Lunch' theorem, which states that all `blind' optimisation algorithms have similar performance when averaged over all conceivable problems, we are optimistic that the Optimal Transport concept may have useful application across a range of inverse problems in geophysics and beyond.

\subsection*{Acknowledgments}
We thank the Commonwealth Scientific Industrial Research Organisation Future Science Platform for Deep Earth Imaging for support. APV acknowledges support from the Australian Research Council through a Discovery Early Career Research Award (grant number DE180100040) and a Discovery Project (grant number DP200100053). This project has received funding from the European Research Council (ERC) under the European Union's Horizon 2020 research and innovation programme grant agreement no. 833848 (UEMHP). The genesis of this work occurred when MS was on a sabbatical visit to the Institute for Geophysics, ETH Z\"{u}rich. The generous support of a Visiting Professorship is gratefully acknowledged. MS would also like to thank Mr. Jared Magyar for conversations that have contributed to his understanding of this subject. The authors are grateful for some constructive reviews, on an earlier version of this manuscript, received from Stan Dosso and Klaus Mosegaard.

\subsection*{Data Availability statement}

Several open source software libraries exist for solution of Optimal Transport problems. One is the {\sl Python Optimal Transport} library of \citet{Flamary:2021}, available from \url{https://pythonot.github.io}, while another is  the {\sl Optimal Transport Tools} of \citet{Cuturi:2022}, available from \url{https://github.com/ott-jax/ot}t, both of which are in Python. The authors' Optimal Transport Python library implementing calculations in this study is available
from \url{https://github. com/msambridge/waveform-ot}. The synthetic seismogram software used in this study, {\sl pyprop8}, is available from \url{https://github.com/valentineap/pyprop8}. 
\bibliography{ms,apv,ot}

\begin{thebibliography}{65}
\expandafter\ifx\csname natexlab\endcsname\relax\def\natexlab#1{#1}\fi

\bibitem[Ambrosio(2003)]{Ambrosio:2003}
Ambrosio, L., 2003.
\newblock {\it Lecture Notes on Optimal Transport Problems\/}, pp. 1--52,
  Springer Berlin Heidelberg, Berlin, Heidelberg.

\bibitem[Benamou \& Brenier(2000)]{Benamou:2000}
Benamou, J. \& Brenier, Y., 2000.
\newblock A computational fluid mechanics solution to the monge-kantorovich
  mass transfer problem, {\it Numer. Math.\/}, {\bf 84}, 375--393,
  doi:10.1007/s002110050002.

\bibitem[Bertsekas \& Castanon(1989)]{Bertsekas:1989}
Bertsekas, D. \& Castanon, D., 1989.
\newblock The auction algorithm for the transportation problem, {\it {Ann.
  Oper. Res.}\/}, {\bf 20}, 67--96.

\bibitem[Bozdağ et~al.(2011)Bozdağ, Trampert, \& Tromp]{Bozdag:2011}
Bozdağ, E., Trampert, J., \& Tromp, J., 2011.
\newblock {Misfit functions for full waveform inversion based on instantaneous
  phase and envelope measurements}, {\it Geophysical Journal International\/},
  {\bf 185}(2), 845--870.

\bibitem[Bracewell(1965)]{Bracewell:1965}
Bracewell, R., 1965.
\newblock {\it The Fourier transformation and its applications\/}, McGraw-Hill,
  New York.

\bibitem[Brenier(1991)]{Brenier:1991}
Brenier, Y., 1991.
\newblock Polar factorization and monotone rearrangement of vector-valued
  functions, {\it Communications on Pure and Applied Mathematics\/}, {\bf
  44}(4), 375--417.

\bibitem[Byrd et~al.(1995)Byrd, Lu, Nocedal, \& Zhu]{Byrd:1995}
Byrd, R., Lu, P., Nocedal, J., \& Zhu, C., 1995.
\newblock A limited memory algorithm for bound constrained optimization, {\it
  SIAM Journal on Scientific Computing\/}, {\bf 16}(5), 1190--1208.

\bibitem[Cuturi(2013)]{Cuturi:2013}
Cuturi, M., 2013.
\newblock Sinkhorn distances: Lightspeed computation of optimal transport, in
  {\em Advances in Neural Information Processing Systems 26\/}, pp. 2292--2300,
  eds Burges, C. J.~C., Bottou, L., Welling, M., Ghahramani, Z., \& Weinberger,
  K.~Q., Curran Associates, Inc.

\bibitem[Cuturi et~al.(2022)Cuturi, Meng-Papaxanthos, Tian, Bunne, Davis, \&
  Teboul]{Cuturi:2022}
Cuturi, M., Meng-Papaxanthos, L., Tian, Y., Bunne, C., Davis, G., \& Teboul,
  O., 2022.
\newblock {Optimal Transport Tools (OTT): A JAX Toolbox for all things
  Wasserstein}, {\it arXiv preprint arXiv:2201.12324\/}.

\bibitem[De~Philippis \& Figalli(2014)]{philippis:2013}
De~Philippis, G. \& Figalli, A., 2014.
\newblock The monge-ampère equation and its link to optimal transportation,
  {\it Bulletin of the American Mathematical Society\/}, {\bf 51}(4), 527 –
  580.

\bibitem[Donno et~al.(2013)Donno, Chauris, \& Calandra]{Donno:2013}
Donno, D., Chauris, H., \& Calandra, H., 2013.
\newblock Estimating the background velocity model with the normalized
  integration method, in {\em 75th Annual International Conference and
  Exhibition\/}, Tu0704, EAGE, Extended Abstracts.

\bibitem[El~Moselhy \& Marzouk(2012)]{ElMoselhy:2012}
El~Moselhy, T.~A. \& Marzouk, Y.~M., 2012.
\newblock Bayesian inference with optimal maps, {\it Journal of Computational
  Physics\/}, {\bf 231}(23), 7815 -- 7850.

\bibitem[Engquist \& Froese(2014)]{Engquist:2014}
Engquist, B. \& Froese, B., 2014.
\newblock Application of the {W}asserstein metric to seismic signals, {\it
  Communications in Mathematical Sciences\/}, {\bf 12}(5), 979--988.

\bibitem[Engquist et~al.(2016)Engquist, Hamfeldt, \& Yang]{Engquist_etal:2016}
Engquist, B., Hamfeldt, B., \& Yang, Y., 2016.
\newblock Optimal transport for seismic full waveform inversion, {\it
  Communications in Mathematical Sciences\/}, {\bf 14}.

\bibitem[Engquist et~al.(2020)Engquist, Ren, \& Yang]{Engquist:2020}
Engquist, B., Ren, K., \& Yang, Y., 2020.
\newblock The quadratic wasserstein metric for inverse data matching, {\it
  Inverse Problems\/}, {\bf 36}(5), 055001.

\bibitem[Fichtner et~al.(2008)Fichtner, Kennett, Igel, \& Bunge]{Fichtner:2008}
Fichtner, A., Kennett, B. L.~N., Igel, H., \& Bunge, H.-P., 2008.
\newblock {Theoretical background for continental- and global-scale
  full-waveform inversion in the time–frequency domain}, {\it Geophysical
  Journal International\/}, {\bf 175}(2), 665--685.

\bibitem[Flamary et~al.(2021)Flamary, Courty, Gramfort, Alaya, Boisbunon,
  Chambon, Chapel, Corenflos, Fatras, Fournier, Gautheron, Gayraud, Janati,
  Rakotomamonjy, Redko, Rolet, Schutz, Seguy, Sutherland, Tavenard, Tong, \&
  Vayer]{Flamary:2021}
Flamary, R., Courty, N., Gramfort, A., Alaya, M.~Z., Boisbunon, A., Chambon,
  S., Chapel, L., Corenflos, A., Fatras, K., Fournier, N., Gautheron, L.,
  Gayraud, N.~T., Janati, H., Rakotomamonjy, A., Redko, I., Rolet, A., Schutz,
  A., Seguy, V., Sutherland, D.~J., Tavenard, R., Tong, A., \& Vayer, T., 2021.
\newblock {POT: Python Optimal Transport}, {\it Journal of Machine Learning
  Research\/}, {\bf 22}(78), 1--8.

\bibitem[Gauthier et~al.(1986)Gauthier, Virieux, \& Tarantola]{Gauthier:1986}
Gauthier, O., Virieux, J., \& Tarantola, A., 1986.
\newblock Two-dimensional nonlinear inversion of seismic waveforms: Numerical
  results, {\it Geophysics\/}, {\bf 51}(7), 1387--1403.

\bibitem[Górszczyk et~al.(2021)Górszczyk, Brossier, \&
  M{\'{e}}tivier]{Gorszczyk:2021}
Górszczyk, A., Brossier, R., \& M{\'{e}}tivier, L., 2021.
\newblock Graph-space optimal transport concept for time-domain full-waveform
  inversion of ocean-bottom seismometer data: Nankai trough velocity structure
  reconstructed from a 1d model, {\it Journal of Geophysical Research: Solid
  Earth\/}, {\bf 126}(5), e2020JB021504, e2020JB021504 2020JB021504.

\bibitem[He et~al.(2019)He, Brossier, M{\'{e}}tivier, \& Plessix]{He:2019}
He, W., Brossier, R., M{\'{e}}tivier, L., \& Plessix, R.-E., 2019.
\newblock {Land seismic multiparameter full waveform inversion in elastic VTI
  media by simultaneously interpreting body waves and surface waves with an
  optimal transport based objective function}, {\it Geophysical Journal
  International\/}, {\bf 219}(3), 1970--1988.

\bibitem[Hedjazian et~al.(2019)Hedjazian, Bodin, \&
  M{\'{e}}tivier]{Hedjazian:2019}
Hedjazian, N., Bodin, T., \& M{\'{e}}tivier, L., 2019.
\newblock An optimal transport approach to linearized inversion of receiver
  functions, {\it Geophysical Journal International\/}, {\bf 216}, 130--147.

\bibitem[Huang et~al.(2019)Huang, Zhang, \& Qian]{Huang:2019}
Huang, G., Zhang, X., \& Qian, J., 2019.
\newblock Kantorovich-{Rubinstein} misfit for inverting gravity-gradient data
  by the level-set method, {\it Geophysics\/}, {\bf 84}, 1--115.

\bibitem[Kantorovich(1942)]{Kantorovich:1942}
Kantorovich, L.~V., 1942.
\newblock On translocation of masses, {\it Dokl. Acad. Nauk. USSR\/}, {\bf 37},
  7--8,227--229 (in Russian). [ English translation: {\it J. Math. Sci.}, vol
  133, 4 (2006), 1381--1382.].

\bibitem[Karmarkar(1984)]{Karmarkar:1984}
Karmarkar, N., 1984.
\newblock A new polynomial-time algorithm for linear programming, {\it
  Combinatorica\/}, {\bf 4}, 373--395.

\bibitem[K\"{a}ufl et~al.(2016)K\"{a}ufl, Valentine, de~Wit, \&
  Trampert]{Kaufl2016a}
K\"{a}ufl, P., Valentine, A., de~Wit, R., \& Trampert, J., 2016.
\newblock Solving probabilistic inverse problems rapidly with prior samples,
  {\it Geophysical Journal International\/}, {\bf 205}, 1710--1728.

\bibitem[Kobayashi et~al.(2006)Kobayashi, Miyazaki, \& Koketsu]{Kobayashi:2006}
Kobayashi, R., Miyazaki, S., \& Koketsu, K., 2006.
\newblock Source processes of the 2005 west off fukuoka prefecture earthquake
  and its largest aftershock inferred from strong motion and 1-hz gps data,
  {\it Earth Planets and Space\/}, {\bf 58}, 57 -- 62.

\bibitem[Kolouri et~al.(2017)Kolouri, Park, Thorpe, Slep\v{c}ev, \&
  Rohde]{Kolouri:2017}
Kolouri, S., Park, S.~R., Thorpe, M., Slep\v{c}ev, D., \& Rohde, G.~K., 2017.
\newblock Optimal mass transport: Signal processing and machine-learning
  applications, {\it IEEE Signal Processing Magazine\/}, {\bf 34}(4), 43--59.

\bibitem[Lellmann et~al.(2014)Lellmann, Lorenz, Schönlieb, \&
  Valkonen]{Lellmann:2014}
Lellmann, J., Lorenz, D.~A., Schönlieb, C., \& Valkonen, T., 2014.
\newblock Imaging with {Kantorovich-Rubinstein} discrepancy, {\it SIAM J. on
  Imag. Sci\/}, {\bf 7 (4)}, 2833--2859.

\bibitem[Levy \& Schwindt(2018)]{Levy:2018}
Levy, B. \& Schwindt, E., 2018.
\newblock Notions of optimal transport theory and how to implement them on a
  computer, {\it Computers and Graphics\/}, {\bf 72}, 135--148.

\bibitem[Luo \& Schuster(1991)]{Luo:1991}
Luo, Y. \& Schuster, G.~T., 1991.
\newblock Wave‐equation traveltime inversion, {\it Geophysics\/}, {\bf
  56}(5), 645--653.

\bibitem[Mainini(2012)]{Mainini:2012}
Mainini, E., 2012.
\newblock A description of transport cost for signed measures, {\it J. Math.
  Sci.\/}, {\bf 181(6)}, 837--855.

\bibitem[M{\'{e}}tivier et~al.(2016{\natexlab{a}})M{\'{e}}tivier, Brossier,
  M{\'{e}}rigot, Oudet, \& Virieux]{Metivier:2016c}
M{\'{e}}tivier, L., Brossier, R., M{\'{e}}rigot, Q., Oudet, e., \& Virieux, J.,
  2016{\natexlab{a}}.
\newblock An optimal transport approach for seismic tomography: application to
  3d full waveform inversion, {\it Inverse Problems\/}, {\bf 32}(11), 115008.

\bibitem[M{\'{e}}tivier et~al.(2016{\natexlab{b}})M{\'{e}}tivier, Brossier,
  Mérigot, Oudet, \& Virieux]{Metivier:2016a}
M{\'{e}}tivier, L., Brossier, R., Mérigot, Q., Oudet, E., \& Virieux, J.,
  2016{\natexlab{b}}.
\newblock Increasing the robustness and applicability of full-waveform
  inversion: An optimal transport distance strategy, {\it The Leading Edge\/},
  {\bf 35}, 1060--1067.

\bibitem[M{\'{e}}tivier et~al.(2016{\natexlab{c}})M{\'{e}}tivier, Brossier,
  Mérigot, Oudet, \& Virieux]{Metivier:2016b}
M{\'{e}}tivier, L., Brossier, R., Mérigot, Q., Oudet, E., \& Virieux, J.,
  2016{\natexlab{c}}.
\newblock {Measuring the misfit between seismograms using an optimal transport
  distance: application to full waveform inversion}, {\it Geophysical Journal
  International\/}, {\bf 205}(1), 345--377.

\bibitem[M{\'{e}}tivier et~al.(2016{\natexlab{d}})M{\'{e}}tivier, Brossier,
  Oudet, Mérigot, \& Virieux]{Metivier:2016d}
M{\'{e}}tivier, L., Brossier, R., Oudet, E., Mérigot, Q., \& Virieux, J.,
  2016{\natexlab{d}}.
\newblock An optimal transport distance for full-waveform inversion:
  Application to the 2014 chevron benchmark data set, {\it SEG Technical
  Program Expanded Abstracts:\/}, pp. 1278--1283.

\bibitem[M{\'{e}}tivier et~al.(2018)M{\'{e}}tivier, Allain, Brossier, Mérigot,
  Oudet, \& Virieux]{Metivier:2018}
M{\'{e}}tivier, L., Allain, A., Brossier, R., Mérigot, Q., Oudet, E., \&
  Virieux, J., 2018.
\newblock Optimal transport for mitigating cycle skipping in full-waveform
  inversion: A graph-space transform approach, {\it Geophysics\/}, {\bf 83},
  R515--R540.

\bibitem[M{\'{e}}tivier et~al.(2019)M{\'{e}}tivier, Brossier, M{\'{e}}rigot, \&
  Oudet]{Metivier:2019}
M{\'{e}}tivier, L., Brossier, R., M{\'{e}}rigot, Q., \& Oudet, E., 2019.
\newblock A graph space optimal transport distance as a generalization of $l_p$
  distances: application to a seismic imaging inverse problem, {\it Inverse
  Problems\/}, {\bf 35}(8), 085001.

\bibitem[Monge(1781)]{Monge:1781}
Monge, G., 1781.
\newblock {\it M{\'e}moire sur la th{\'e}orie des d{\'e}blais et des
  remblais\/}, De l'Imprimerie Royale.

\bibitem[Okazaki et~al.(2021)Okazaki, Hachiya, Iwaki, Maeda, Fujiwara, \&
  Ueda]{Okazaki:2021}
Okazaki, T., Hachiya, H., Iwaki, A., Maeda, T., Fujiwara, H., \& Ueda, N.,
  2021.
\newblock {Simulation of broad-band ground motions with consistent long-period
  and short-period components using the Wasserstein interpolation of
  acceleration envelopes}, {\it Geophysical Journal International\/}, {\bf
  227}(1), 333--349.

\bibitem[O'Toole \& Woodhouse(2011)]{OToole2011}
O'Toole, T. \& Woodhouse, J., 2011.
\newblock Numerically stable computation of complete synthetic seismograms
  including the static displacement in plane layered media, {\it Geophysical
  Journal International\/}, {\bf 187}, 1516--1536.

\bibitem[O'Toole et~al.(2012)O'Toole, Valentine, \& Woodhouse]{OToole2012}
O'Toole, T., Valentine, A., \& Woodhouse, J., 2012.
\newblock Centroid--moment tensor inversions using high-rate {GPS} waveforms,
  {\it Geophysical Journal International\/}, {\bf 191}, 257--270.

\bibitem[Peyr\'e \& Cuturi(2019)]{Peyre:2019}
Peyr\'e, G. \& Cuturi, M., 2019.
\newblock Computational optimal transport, {\it Foundations and Trends in
  Machine Learning\/}, {\bf 11}(5-6), 355--607.

\bibitem[Pladys et~al.(2021)Pladys, Brossier, Youbing, \&
  M{\'{e}}tivier]{Pladys:2021}
Pladys, A., Brossier, R., Youbing, L., \& M{\'{e}}tivier, L., 2021.
\newblock {On cycle-skipping and misfit function modification for full-wave
  inversion: Comparison of five recent approaches}, {\it Geophysics\/}, {\bf
  86}(4), R563--R587.

\bibitem[Rawlinson \& Sambridge(2004)]{RawlinsonA:2002}
Rawlinson, N. \& Sambridge, M., 2004.
\newblock Wavefront evolution in strongly heterogeneous layered media using the
  {F}ast {M}arching {M}ethod, {\it Geophys. J. Int.\/}, {\bf 156}, 631--647,
  doi: 10.1111/j.1365--246X.2004.02153.x.

\bibitem[Santambrogio(2015)]{Santambrogio:2015}
Santambrogio, F., 2015.
\newblock {\it Optimal Transport for Applied Mathematicians. Calculus of
  Variations, PDEs and Modeling\/}, Birkhäuser.

\bibitem[Sethian(1996)]{Sethian:1996}
Sethian, J.~A., 1996.
\newblock A fast marching level set method for monotonically advancing fronts,
  {\it Proc. Nat. Acad. Sci.\/}, {\bf 93}, 1591--1595.

\bibitem[Sethian(1999)]{SethianA:1999}
Sethian, J.~A., 1999.
\newblock {\it Level Set Methods and Fast Marching Methods\/}, Cambridge
  University Press, Cambridge.

\bibitem[Sethian \& Popovici(1999)]{Sethian:1999}
Sethian, J.~A. \& Popovici, A.~M., 1999.
\newblock 3-{D} traveltime computation using the fast marching method, {\it
  Geophysics\/}, {\bf 64}, 516--523.

\bibitem[Sieminski et~al.(2007)Sieminski, Liu, Trampert, \&
  Tromp]{Sieminski:2006}
Sieminski, A., Liu, Q., Trampert, J., \& Tromp, J., 2007.
\newblock Finite-frequency sensitivity of surface waves to anisotropy based on
  adjoint methods, {\it Geophys. J. Int.\/}, {\bf 168}, 1153--1174.

\bibitem[Sirgue \& Pratt(2004)]{sirgue:2004}
Sirgue, L. \& Pratt, R.~G., 2004.
\newblock Efficient waveform inversion and imaging: A strategy for selecting
  temporal frequencies, {\it Geophysics\/}, {\bf 69}(1), 231--248.

\bibitem[Solomon(2015)]{Solomon:2015a}
Solomon, J., 2015.
\newblock {\it Transportation Techniques for Geometric Data Processing\/},
  Ph.D. thesis, Stanford Department of Computer Science.

\bibitem[Solomon et~al.(2015)Solomon, De~Goes, Peyré, Cuturi, Butscher,
  Nguyen, Du, \& Guibas]{Solomon:2015}
Solomon, J., De~Goes, F., Peyré, G., Cuturi, M., Butscher, A., Nguyen, A., Du,
  T., \& Guibas, L., 2015.
\newblock Convolutional {W}asserstein distances: Efficient optimal
  transportation on geometric domains, {\it {ACM Transactions on Graphics}\/},
  {\bf 34}(4), 66:1--66:11.

\bibitem[Sun \& Alkhalifah(2019)]{Sun:2019}
Sun, B. \& Alkhalifah, T., 2019.
\newblock {The application of an optimal transport to a preconditioned data
  matching function for robust waveform inversion}, {\it Geophysics\/}, {\bf
  84}(6), R923--R945.

\bibitem[Taner et~al.(1979)Taner, Koehler, \& Sheriff]{Taner:1979}
Taner, M.~T., Koehler, F., \& Sheriff, R., 1979.
\newblock Complex seismic trace analysis, {\it Geophysics\/}, {\bf 44}(6),
  1041--1063.

\bibitem[Tromp et~al.(2005)Tromp, Tape, \& Liu]{Tromp:2005}
Tromp, J., Tape, C., \& Liu, Q., 2005.
\newblock Seismic tomography, adjoint methods, time reversal, and banana-donut
  kernels, {\it Geophys. J. Int.\/}, {\bf 160}, 195--216, doi:
  10.1111/j.1365--246X.2004.02456.x.

\bibitem[Valentine \& Sambridge(2021)]{Valentine:2021}
Valentine, A.~P. \& Sambridge, M., 2021.
\newblock pyprop8: A lightweight code to simulate seismic observables in a
  layered half-space, {\it Journal of Open Source Software\/}, {\bf 6}(66),
  3858.

\bibitem[Villani(2003)]{Villani:2003}
Villani, C., 2003.
\newblock {\it Topics in Optimal Transportation\/}, Graduate studies in
  mathematics, American Mathematical Society.

\bibitem[Villani(2008)]{Villani:2008}
Villani, C., 2008.
\newblock {\it Optimal Transport: Old and New\/}, Grundlehren der
  mathematischen Wissenschaften, Springer Berlin Heidelberg.

\bibitem[Virieux \& Operto(2009)]{Virieux:2009}
Virieux, J. \& Operto, S., 2009.
\newblock An overview of full-waveform inversion in exploration geophysics,
  {\it Geophysics\/}, {\bf 74}(6), WCC1--WCC26.

\bibitem[Virtanen et~al.(2020)Virtanen, Gommers, Oliphant, Haberland, Reddy,
  Cournapeau, Burovski, Peterson, Weckesser, Bright, {van der Walt}, Brett,
  Wilson, Millman, Mayorov, Nelson, Jones, Kern, Larson, Carey, Polat, Feng,
  Moore, {VanderPlas}, Laxalde, Perktold, Cimrman, Henriksen, Quintero, Harris,
  Archibald, Ribeiro, Pedregosa, {van Mulbregt}, \& {SciPy 1.0
  Contributors}]{Virtanen:2020}
Virtanen, P., Gommers, R., Oliphant, T.~E., Haberland, M., Reddy, T.,
  Cournapeau, D., Burovski, E., Peterson, P., Weckesser, W., Bright, J., {van
  der Walt}, S.~J., Brett, M., Wilson, J., Millman, K.~J., Mayorov, N., Nelson,
  A. R.~J., Jones, E., Kern, R., Larson, E., Carey, C.~J., Polat, {\.I}., Feng,
  Y., Moore, E.~W., {VanderPlas}, J., Laxalde, D., Perktold, J., Cimrman, R.,
  Henriksen, I., Quintero, E.~A., Harris, C.~R., Archibald, A.~M., Ribeiro,
  A.~H., Pedregosa, F., {van Mulbregt}, P., \& {SciPy 1.0 Contributors}, 2020.
\newblock {{SciPy} 1.0: Fundamental Algorithms for Scientific Computing in
  Python}, {\it Nature Methods\/}, {\bf 17}, 261--272.

\bibitem[Warner \& Guasch(2016)]{Warner:2016}
Warner, M. \& Guasch, L., 2016.
\newblock Adaptive waveform inversion: Theory, {\it Geophysics\/}, {\bf 81},
  R429--R445.

\bibitem[Wu et~al.(2014)Wu, Luo, \& Wu]{Wu:2014}
Wu, R.-S., Luo, J., \& Wu, B., 2014.
\newblock Seismic envelope inversion and modulation signal model, {\it
  Geophysics\/}, {\bf 79}(3), WA13--WA24.

\bibitem[Yang \& Engquist(2018)]{Yang:2018}
Yang, Y. \& Engquist, B., 2018.
\newblock {Analysis of optimal transport and related misfit functions in
  full-waveform inversion}, {\it Geophysics\/}, {\bf 83}(1), A7--A12.

\bibitem[Yang et~al.(2018)Yang, Engquist, Sun, \& Hamfeldt]{Yang_etal:2018}
Yang, Y., Engquist, B., Sun, J., \& Hamfeldt, B.~F., 2018.
\newblock {Application of optimal transport and the quadratic Wasserstein
  metric to full-waveform inversion }, {\it Geophysics\/}, {\bf 83}(1),
  R43--R62.

\bibitem[Zhu et~al.(1997)Zhu, Byrd, Lu, \& Nocedal]{Zhu:1997}
Zhu, C., Byrd, R., Lu, P., \& Nocedal, J., 1997.
\newblock Algorithm 778: {L-BFGS-B: Fortran Subroutines for Large-Scale
  Bound-Constrained Optimization}, {\it ACM Transactions on Mathematical
  Software\/}, {\bf 23}(4), 550--560.

\end{thebibliography}
\bibliographystyle{gji}

\newpage
\appendix

\section{Optimal Transport by Linear Programming}\label{app:LP}

\cite{Kantorovich:1942}  recast Optimal Transport as a Linear Programming constrained optimization problem. The discrete point mass representation of $f(x)$ and $g(y)$ in (\ref{eq:pointmass}) means that the transport plan can be represented by the matrix $\pi_{ij}, (i=1,\dots, n_g; j=1\dots, n_f)$,  where the transport from $f$ to $g$ is given by the matrix vector product in (\ref{eq:discreteT0})
\begin{equation}
{\bf g} = {\mathbf{\pi}} {\bf f},
\label{eq:discreteT}
\end{equation}
where ${\bf f} = (f_1,f_2,\dots,n_f)^T$ and ${\bf g} = (g_1,g_2,\dots,n_g)^T$. If $c_{ij}$ is the cost of transporting $f_i$ to $g_j$ then the total cost of transporting $\mathbf{f}$ to $\mathbf{g}$ with the transport plan $\mathbf{\pi}$ is
\begin{equation}
I = \sum_i \sum_j \pi_{ij} c_{ij}.
\label{eq:discreteI}
\end{equation}
The Optimal Transport problem is then to seek the plan, $ {\mathbf{\pi}}$, which minimizes the total work (\ref{eq:discreteI}). As \cite{Kantorovich:1942} pointed out, this corresponds to the following Linear Programming problem
\begin{equation}
W_p^p = \min_{\pi_{ij}} \sum_i \sum_j \pi_{ij} c_{ij}.
\label{eq:discreteW}
\end{equation}
subject to row and column sums
\begin{equation}
\quad  \sum_j \pi_{ij} = f_i, ,\quad (i=1,\dots,n_f)
\label{eq:discretesumf}
\end{equation}
and
\begin{equation}
 \quad \sum_i \pi_{ij} = g_j,,\quad (j=1,\dots,n_g).
\label{eq:discretesumg}
\end{equation}
The type of distance, e.g. absolute difference or Euclidean, is a choice and defines the nature of the Optimal Transport process and characteristics of $W_p^p$ as seen in Figs. \ref{fig:ricker_align} and  \ref{fig:rickermisfit}. Since probabilities must be positive the inequality constraints
\begin{equation}
\quad  \pi_{ij} \ge 0, \quad (i=1,\dots,n_i;j=1,\dots,n_g),
\label{eq:pos}
\end{equation}
must also be imposed.

As noted in the main text, $\mathbf{\pi}$ corresponds to a density function in the product space of $\mathbf f$ and $\mathbf g$, with each becoming one of its marginals, as illustrated in Fig.  \ref{fig:tplans}.
As with all Linear Programming problems the solution must lie at the boundary of the (in)equality constraints (\ref{eq:discretesumf})-(\ref{eq:pos}). Once $\mathbf{\pi}$ is known then the corresponding $p$-Wasserstein distance is given by (\ref{eq:discreteW}), while (\ref{eq:discreteT}) transforms $\mathbf{f}$ onto $\mathbf{g}$. Modern techniques of solution, e.g. interior point methods \citep{Karmarkar:1984} may be used, however with a computational cost that typically scales with the cube of number of unknowns \citep{Solomon:2015}. In most cases this will be prohibitive. Nevertheless this is the only approach to Wasserstein distance calculation that applies to any cost function, any $p$ value, and for densities $f(x)$ and $g(y)$ in any number of dimensions.

\newpage\section{Derivatives of 1D Wasserstein distance}\label{app:Wderiv}

\subsection{Derivatives of the Wasserstein distance with respect to point mass amplitudes} 
In an inversion context where the Wasserstein distance  (\ref{eq:Wp}) is used as a misfit function between observations and model predictions, we will assume that it is the source density, $f(x)$, which is associated with the predictions and the target density,  $g(y)$, with observations.  Hence the derivatives of $W^p_p$ with respect to source amplitudes are required. Here we derive expressions for these derivatives through differentiation of  (\ref{eq:Wp}).
By definition the source density amplitudes are normalised, which we write as
\begin{equation}
f_i=  \frac{ \tilde f_i}{\sum_k \tilde f_k}, \quad (i=1,\dots,n_f),
\label{eq:derivfi}
\end{equation}
where the tilde indicates an unnormalized density. 
Depending on the application, derivatives with respect to normalized, $\frac{\partial W^p_p}{\partial f_i}, (i=1,\dots,n_f)$,  or unnormalised, $\frac{\partial W^p_p}{\partial \tilde f_i}, (i=1,\dots,n_f)$,  density amplitudes will be required. 

Recall that the variables $t_k, (k=1,\dots,k_{max})$ in section (\ref{subsec:dpm}) represent the piecewise constant intervals of the cumulative distribution functions of $F(x)$ and $G(y)$. Since these only depend on the amplitudes, $f_i$, and not the location of point masses, $z_k$, then from (\ref{eq:Wp}) we have
\begin{equation} 
\frac{\partial W_p^p(f,g)}{\partial \tilde f_i} = \Delta {\bf  z}^T A {\bf t}^{\prime},
\label{eq:derivfj3}
\end{equation}
where 
\begin{equation} 
{\bf t}^{\prime} = \left [  \frac{\partial t_1}{\partial \tilde f_i}, \frac{\partial t_2}{\partial\tilde  f_i},\dots, \frac{\partial t_{k_{max}}}{\partial \tilde f_i} \right ]^T .
\label{eq:derivfj4}
\end{equation}
Clearly all of the derivatives of cumulative variables, $t_k$, which correspond to the target, $G(y)$, will be zero, and so we write
\begin{equation} 
  \frac{\partial t_k}{\partial \tilde f_i} = q_{k} \frac{\partial F_{i(k)}}{\partial \tilde f_i}, \quad (i=1,\dots,n_g;k=1,\dots,k_{max}),
\label{eq:derivfj5}
\end{equation}
where $q_{k} = 1$ if $t_k$ lies in the $F$ set, and zero otherwise. The terms $F_{i(k)}$ represent the value of the cumulative distribution term corresponding to $t_k$.

To complete the derivation, from (\ref{eq:derivfi}) we have
\begin{eqnarray} 
\frac{\partial f_j}{\partial \tilde f_i}&=& \frac{1-f_i}{\sum_k \tilde f_k}, \quad (i=j), \label{eq:derivfja}\\
&=& \frac{-f_j}{\sum_k \tilde f_k}, \quad (i\ne j).
\label{eq:derivfjb}
\end{eqnarray}
Using (\ref{eq:cumul1Df}), together with (\ref{eq:derivfja}) - (\ref{eq:derivfjb}) gives
\begin{equation} 
\frac{\partial F_i(k)}{\partial \tilde f_i} = \frac{1}{\sum_k \tilde f_k} \left [ h_{i,i(k)} - F_{i(k)} \right ].
\label{eq:derivfinal}
\end{equation}
where $h_{ij} = 1,$ if $ i \le j$ and $h_{ij} = 0,$ if $ i>j$. 
Combining (\ref{eq:derivfinal}),  (\ref{eq:derivfj5}) in (\ref{eq:derivfj4}) allows calculation of all $n_f$ derivatives of the $p$-Wasserstein distance with respect to unnormalized amplitudes of $f(x)$. Here we have assumed that each of the values of $t_k$ are distinct.

For normalised derivatives we note that, due to similar arguments, expressions (\ref{eq:derivfj3}) - (\ref{eq:derivfj5}) may be replaced by their normalized counterparts, i.e. (\ref{eq:derivfj5}) becomes.
\begin{equation} 
  \frac{\partial t_k}{\partial f_i} = q_{k} \frac{\partial F_{i(k)}}{\partial f_i}, \quad (i=1,\dots,n_g;k=1,\dots,k_{max}),
\label{eq:derivfj5n}.
\end{equation}
Then from the definition of the cumulative distribution function for $F(x)$ in (\ref{eq:cumul1Df}) we have
\begin{equation} 
  \frac{\partial t_k}{\partial f_i} = q_{k} h_{i,i(k)}, \quad (i=1,\dots,n_g;k=1,\dots,k_{max}).
\label{eq:derivfj5n2}
\end{equation}

\newpage
\section{Calculating the seismogram fingerprint and its derivatives}\label{app:FP}
The seismogram fingerprint  density, Figs. \ref{fig:rickerfp}  and \ref{fig:FPandPDF}, has two steps in its calculation. Given a time-amplitude window $(\Delta_u,\Delta_t)$, the first step is to find the nearest distance, $d_{ij}$, for each point on a regular grid, ${\bf p}_{ij}, (i=1,\dots,n_t;j=1,\dots,n_u)$ to the waveform, $u(t)$, which is represented as a series of piecewise linear segments, ${\bf x}_k = [t_k, u_k]^T, (k=1,\dots,n)$, on the $(t,u)$ plane (see Fig. \ref{fig:segments}).  The second is then to calculate the unnormalised 2D  density function at each point on the grid
\begin{equation}
p_{ij} =e^{-d_{ij}/s} \quad(i=1,\dots,n_t; j=1,\dots,n_u),
\label{eq:PDF}
 \end{equation}
where $s$ is a distance scale factor. In principle we might also choose to replace the distance with its square, but have found no compelling reason to do so in this work. In this appendix we provide details of a simple algorithm for calculation of the distance field as well as the derivatives of the  density function amplitudes, $p_{ij}$, with respect to the waveform amplitudes, $u_k = u (t_k)$. 

\subsection{Nearest distance of grid points to waveform}
In calculating the distance field, $d_{ij}$, we face the choice of whether  to represent the waveform as a series of discrete points only, ${\bf x}_k, (k=1,\dots,n)$, or more accurately as piecewise linear segments between those points (see Fig.  \ref{fig:segments}). In the first case the nearest distance field, $d_{ij}$, can be efficiently found with the Fast Marching algorithm of \citet{Sethian:1996} \citep[see also] []{ Sethian:1999, RawlinsonA:2002}. This approach has the advantage that its computation time depends only on the size of the $n_t \times n_u$ grid, and not on the number of segments in the waveform, $n$. However, a drawback is the fact that the distance field shape becomes distorted close to the waveform (see the pink contours in Fig. \ref{fig:segments}). Therefore we take the second option and represent the waveform as a series of piecewise linear segments, which better resembles the character of the waveform at close distances (see the blue contours in Fig. \ref{fig:segments}). 

The task then is to locate the segment containing the nearest waveform point to each grid node. An example in Fig.  \ref{fig:segments} is the point marked ${\bf x}(\lambda^*)$ which is closest to ${\bf p}_{ij}$, and lies in the segment between ${\bf x}_k$ and ${\bf x}_{k+1}$. Unfortunately, there appears to be no analogue of the Fast Marching method for this case, and so we resort to a simple enumerative search of all segments for each grid node to locate all nearest points and their distances. While the computational costs of enumeration will scale with the product $n_u \times n_t\times  n$ this calculation may be performed in parallel and can usually be made efficient with modern programming languages that can make use of vectorization, or parallelization.

The question remains then of how to calculate the nearest point ${\bf x}(\lambda^*)$ from ${\bf p}_{ij}$ for the $k$th segment. To do this, we first define any point along the $k$th segment by
\begin{equation}
{\bf x}(\lambda) = (1-\lambda) {\bf x}_k + \lambda {\bf x}_{k+1}
\label{eq:xlam}
 \end{equation}
Some elementary geometry leads to the value of $\lambda$ corresponding to the nearest point to ${\bf p}_{ij}$
\begin{equation}
\lambda^* = \frac{({\bf p}_{ij} - {\bf x}_k) . ({\bf x}_{k+1}- {\bf x}_k)}{|| {\bf x}_{k+1}- {\bf x}_k) ||^2}
\label{eq:lamstar}
 \end{equation}
while noting that a value $\lambda^* \le 0$ indicates that the nearest node is ${\bf x}_k$ and conversely $\lambda^* \ge 1$ indicates that it must be ${\bf x}_{k+1}$. The nearest distance from the $(i,j)$th grid node to the $k$-th segment is then clearly
\begin{equation}
 l_{ijk} = || {\bf p}_{ij} - {\bf x}(\lambda^*) ||,
\label{eq:lamdist}
 \end{equation}
and we have
\begin{equation}
d_{ij}=\min_{k} l_{ijk}.
\label{eq:ndist}
 \end{equation}
This calculation can be repeated for each grid node to obtain the distance function on the $(n_t \times n_u)$ grid.

\subsection{Derivatives}
The derivatives required are that of the density function at each grid node, $p_{ij}$, in (\ref{eq:PDF}), with respect to the height of each waveform node, $u_k$. From (\ref{eq:PDF}) we have
\begin{equation}
\frac{\partial p_{ij}}{\partial u_k} =\frac{-p_{ij} }{s} \frac{\partial d_{ij}}{\partial u_k}
\quad (i=1,\dots,n_t;j=1,\dots,n_u; k=1,\dots,n).
\label{eq:PDFderiv}
 \end{equation}
From  Fig. (\ref{fig:segments}) it is clear that for each grid node $(i,j)$, the closest waveform point will either be along a segment, like ${\bf p}$, or at an end point, like ${\bf p}^{\prime}$. In the former case the location of the ${\bf x}(\lambda)$ depends only on the nodes ${\bf x}_k$ and ${\bf x}_{k+1}$, and so it is only derivatives with respect to $u_k$ and $u_{k+1}$ that are non zero. Similarly, in the latter case the only distance function derivative that is non zero is with respect to $u_k$. If we assume we know which segment contains the closest point to the grid node (which we do from the enumerative search), it suffices to consider the case of a single segment and differentiate (\ref{eq:lamdist}) with respect to  $u_k$ and $u_{k+1}$. We have then for closest segment $k$
\begin{eqnarray}
\frac{\partial d_{ij}}{\partial u_k} &=& \frac{\partial d_{ij}}{\partial {\bf x}(\lambda)}. \frac{\partial{\bf x}(\lambda))}{\partial u_k}, \\
&=& \frac{({\bf x}_k-{\bf p}_{ij})}{d_{ij}}. (1-\lambda^*){\bf e}_u \\
&=& (1-\lambda^*)\frac{(u_k-\tilde u_j)}{d_{ij}},
\label{eq:derivk}
 \end{eqnarray}
where ${\bf e}_u$ is the unit vector in the amplitude (vertical) direction of the $(t,u)$ plane.
The keen reader will notice that the second part of the derivative expression is just the sine of the take off angle of the `ray' from the waveform point ${\bf x}(\lambda^*)$ to the grid point ${\bf p}_{ij}$ (i.e. the green lines in Figs. \ref{fig:segments} and \ref{fig:rays}). This is unsurprising because, as noted previously, the calculation of the distance field is identical to that of a seismic travel time field in a unit wavespeed homogeneous medium, with the waveform as an initial wavefront. The lines connecting grid nodes and their closet point on the waveform become `seismic raypaths' in this analogy, and hence the derivatives of the distance field with respect to the location of ${\bf x}(\lambda^*)$ are the same as that for seismic travel times with respect to source location. The coefficient $(1-\lambda^*)$ merely translates this derivative to the waveform amplitude of the $k$th node. Using the same approach we can obtain the derivative of the distance $d_{ij}$ with respect to the amplitude at the $(k+1)$th node
\begin{equation}
\frac{\partial d_{ij}}{\partial u_{k+1}} =\lambda^*\frac{(u_{k+1}-\tilde u_j)}{d_{ij}}.
\label{eq:derivk1}
 \end{equation}
For the case of a grid node whose closest point lies at a segment end point, like ${\bf p}^{\prime}$ in Fig. (\ref{fig:segments}),  (\ref{eq:derivk}) can be used by setting $\lambda^*=0$. This completes the derivative calculations of the unnormalised density function with respect to waveform node amplitudes. Combining the above expressions we have
\begin{eqnarray}
\frac{\partial p_{ij}}{\partial u_k}&=&\frac{-p_{ij}(1-\lambda^*) }{s} \frac{(u_k-\tilde u_j)}{d_{ij}} \quad (i=1,\dots,n_t;j=1,\dots,n_u) \label{eq:derivpu1}\\
\frac{\partial p_{ij}}{\partial u_{k+1}}&=&\frac{-p_{ij}\lambda^* }{s} \frac{(u_{k+1}-\tilde u_j)}{d_{ij}} \quad (i=1,\dots,n_t;j=1,\dots,n_u)\\
&=&0 \quad {\rm otherwise}.
\label{eq:derivpu3}
\end{eqnarray}
where $k$  is the segment containing the closest point to grid node $(i,j)$ as given by (\ref{eq:ndist}), and noting that if the endpoint $k$ is closest, as it will be for many grid points, then we can use the same expressions with $\lambda^*=0$. For completeness we also require the derivatives of the normalised  density amplitudes, $\bar p_{qr} $ and these can then be found from
\begin{equation}
\bar p_{qr} = \frac{p_{qr}}{\sum_{i} \sum_{j} p_{ij}},
\label{eq:pnorm}
 \end{equation}
 which gives
\begin{equation}
\frac{\partial \bar p_{qr}}{\partial p_{ij}} = \frac{(\delta_{iq} \delta_{jr} - \bar p_{qr} )}{\sum_{l} \sum_{m} p_{lm}}
\label{eq:derivPnorm}
 \end{equation}
and once again the normalised derivatives can be calculated in straightforward manner by application of the chain rule.

As noted above, here we use an enumerative search to locate the minimum distance in (\ref{eq:ndist}) which also gives either the nearest segment, or, if required, the nearest end point along the waveform. It is worth noting that because the waveform can have arbitrary shape, these two are not necessarily related, i.e. they nearest end point might not be attached to the nearest segment.  Hence it is not 
in general possible to locate one from the other.  This is a secondary benefit of performing enumerative search in solving (\ref{eq:ndist}), in that exact identification of both nearest waveform segment and waveform point is guaranteed, whereas the Fast Marching algorithm, for example, would only provide the latter and not the former.

In summary, all derivatives of the normalised  density function (\ref{eq:PDFderiv}), with respect to waveform amplitudes are straightforward to implement, involve only evaluation of analytical expressions, and can also utilize vectorization, or parallelisation, features in modern programming languages.
\begin{figure}
\begin{center}
\includegraphics[width=0.5\textwidth]{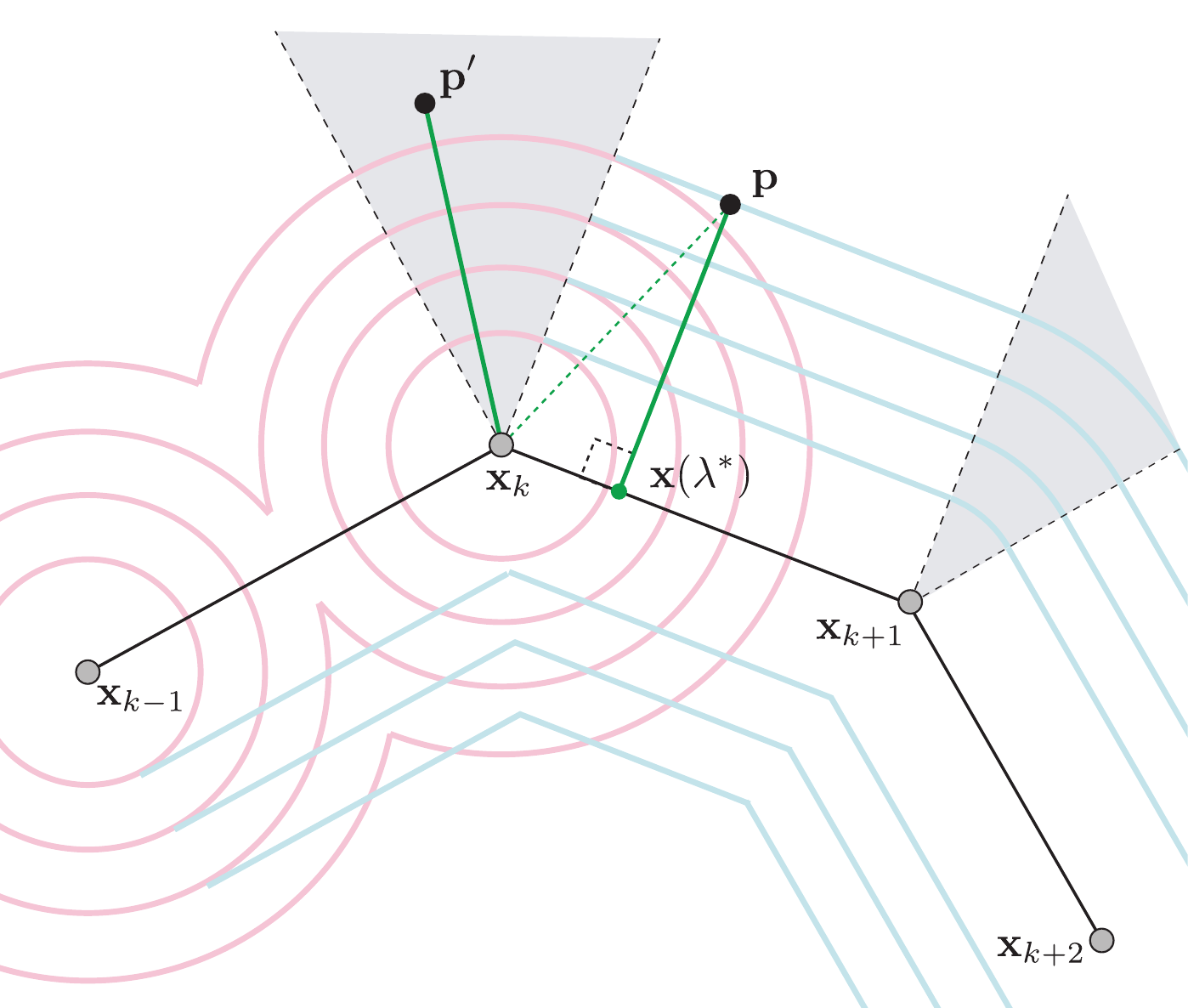}
\end{center}
\caption{\label{fig:segments} Constructing the nearest distance field about a seismic waveform. Here a waveform is represented by a series of linear segments defined between points ${\bf x}_k, (k=1,\dots,n)$, while ${\bf p}$ and ${\bf p}^{\prime}$  are two points on the time-amplitude plane. The latter lies in the shaded area and will be closest to an end point of a waveform segment,  here ${\bf x}_k$, while point ${\bf p}$  lies outside of the shaded areas and will be closest to a point along a waveform segment, here denoted by ${\bf x}(\lambda^*)$. Green lines are the `rays' from ${\bf p}$ and ${\bf p}^{\prime}$ to their closest points on the waveform. Contours are shown of the nearest distance function ignoring ray segments (pink) and including them (blue). }
\end{figure}

\begin{figure}
\begin{center}
\includegraphics[width=0.9\textwidth]{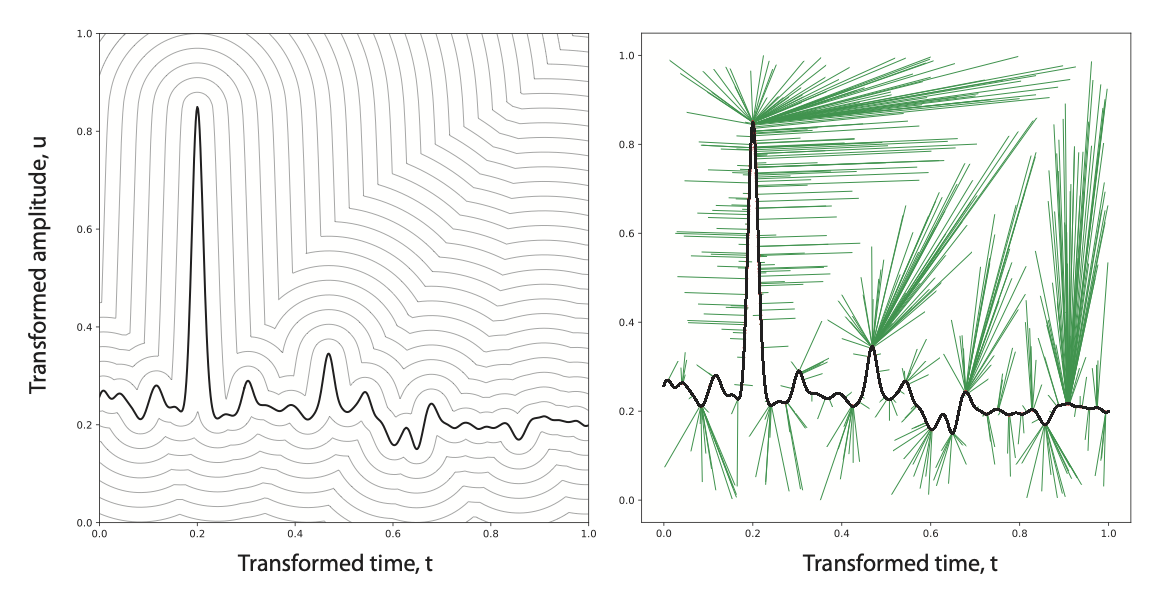}
\end{center}
\caption{\label{fig:rays} The left panel shows contours of the nearest distance field about a synthetic seismic Receiver Function. This resembles a fingerprint and can be used as a 2D image, in place of the underlying time series, for calculation of misfit measures between predicted and observed seismograms. The right panel shows the geometric ray paths from selected points on the plane to their nearest point on the waveforms. These calculations are mathematically equivalent to that of a seismic travel time field and raypaths (see text).}
\end{figure}

\newpage
\section{Marginal Wasserstein and its derivatives}\label{app:margwass}
 
Given the 2D  density function, $\bar p_{ij}$  from (\ref{eq:PDF}) and (\ref{eq:pnorm}), the calculation of the two marginal density functions in Figs. \ref{fig:rickerfp}c \& d, is given by summation over the relevant index. Specifically, we have the 1D marginals for time, $p^{(t)}_i, (i=1,\dots,n_t)$, and amplitude, $p^{(u)}_j, (j=1,\dots,n_u)$  given by
\begin{equation}
p^{(t)}_i = \sum_j \bar p_{ij}, \quad\quad p^{(u)}_j = \sum_i \bar p_{ij},
\label{eq:marg}
 \end{equation}
As described in the main text, the Wasserstein distance between observed and predicted fingerprints is then approximated by the sum of the 1D Wasserstein distances between corresponding marginals, which we write as
\begin{equation}
W_{f} = \alpha W_t + (1-\alpha)W_u,
\label{eq:Wmarg}
 \end{equation}
 where to simplify notation we have dropped the $p$ subscript and note that $W_f$ can represent either $W_1$ or $W_2^2$.
\subsection{Derivatives}

The 1D Wasserstein theory described in section \ref{sec:theory} and appendix \ref{app:Wderiv} provide the derivatives of Wasserstein distance with respect to the amplitude of the input source density.  Specifically for the  case of 1D marginals the derivatives 
$\partial W/ \partial p^{(t)}_i, (i=1,\dots,n_t)$ and $\partial W/ \partial p^{(u)}_j, (j=1,\dots,n_u)$, are found from (\ref{eq:derivfj3})-(\ref{eq:derivfj4}).
We now provide the relationship between these derivatives and those of the Wasserstein distance in (\ref{eq:Wmarg}) with respect to the 2D density amplitudes, $\bar p_{ij}$. From (\ref{eq:Wmarg}) we have
\begin{equation}
\frac{\partial W_{f}}{ \partial \bar p_{ij}} =\alpha \frac{\partial W_t}{ \partial \bar p_{ij}} + (1-\alpha )\frac{\partial W_u}{ \partial \bar p_{ij}} 
\label{eq:Wderiv2D}
 \end{equation}
Since the marginals, $p^{(t)}_i$ and $p^{(u)}_j$, are functions of the 2D amplitudes $\bar p_{ij}$ we can write
\begin{equation}
\frac{\partial W_t}{ \partial \bar p_{ij}} = \frac{\partial W_t}{ \partial  p^{(t)}_i} \frac{\partial p^{(t)}_i}{ \partial \bar p_{ij}},\quad(i=1,\dots,n_t; j=1,\dots,n_u),
\label{eq:Wderivt}
 \end{equation}
and from (\ref{eq:marg}) we see that the second term is unity and so
\begin{equation}
\frac{\partial W_t}{ \partial \bar p_{ij}} =  \frac{\partial W_t}{ \partial  p^{(t)}_i},\quad(i=1,\dots,n_t; j=1,\dots,n_u).
\label{eq:Wderivt2}
 \end{equation}
By a similar argument for the amplitude marginal we get
\begin{equation}
 \frac{\partial W_u}{ \partial \bar p_{ij}} = \frac{\partial W_u}{  \partial p^{(u)}_j},\quad(i=1,\dots,n_t; j=1,\dots,n_u)).
\label{eq:Wderivu}
 \end{equation}
Substituting these expressions into (\ref{eq:Wderiv2D}) we arrive at an expression for the derivatives of the Wasserstein distance with respect to the normalised 2D fingerprint  density amplitudes in terms of the marginal  density amplitudes.
\begin{equation}
\frac{\partial W_{f}}{ \partial \bar p_{ij}} =\alpha \frac{\partial W_t}{\partial  p^{(t)}_i} + (1-\alpha )\frac{\partial W_u}{ \partial p^{(u)}_j}.
\label{eq:Wderiv2Dp}
 \end{equation}
It can be useful, depending on implementation, to also have the derivatives of the Wasserstein distance with respect to the unnormalized  density amplitudes in (\ref{eq:pnorm}). Again from the chain rule we have
\begin{equation}
\frac{\partial W_{f}}{\partial p_{ij}} = \sum_{qr} \frac{\partial W_{f}}{\partial \bar p_{qr}}  \frac{\partial \bar p_{qr}} {\partial p_{ij}}
\label{eq:Wderiv2Dchain}
 \end{equation}
Using  (\ref{eq:pnorm}) and simplifying we obtain
\begin{equation}
\frac{\partial W_{f}}{\partial p_{ij}} = \frac{1}{\sum_{qr} p_{qr}} \left[ \frac{\partial W_{f}}{\partial \bar p_{ij}}  - 
\sum_{qr} \frac{\partial W_{f}}{\partial \bar p_{qr}} \bar p_{qr}.
\right]
\label{eq:Wderiv2Dun}
 \end{equation}
By combining the derivative expressions from appendices \ref{app:FP} and \ref{app:margwass} we  obtain the derivatives of the Wasserstein distance, $W_{f}$, with respect to the amplitudes of the waveform $u_k$. Specifically we have
\begin{equation}
\frac{\partial W_{f}}{\partial u_k} = \sum_{ij} \frac{\partial W_{f}}{\partial p_{ij}} \frac{\partial p_{ij}}{\partial u_k},\quad (k=1,\dots,n),
\label{eq:WderivdU}
 \end{equation}
where the first set of derivatives on the right hand side are given by (\ref{eq:Wderiv2Dun}) and (\ref{eq:Wderiv2Dp}), and the second set by (\ref{eq:derivpu1}) - (\ref{eq:derivpu3}). Recall that each node in the 2D grid of  density amplitudes $p_{ij}$  has a single raypath back to the waveform segment containing its closest point, as shown in Fig. \ref{fig:rays}.  Therefore for each grid node, $(i,j)$, there will be at most two derivatives $\frac{\partial p_{ij}}{\partial u_k}$ that are non zero, i.e. for the two waveform points $k$ and $k+1$ defining the waveform segment containing the ray from $(i,j)$, see Fig. \ref{fig:segments}. Therefore the matrix of size $(n_tn_u \times n)$ in (\ref{eq:WderivdU}) is highly sparse and the evaluation of the summation may be performed efficiently by only considering non zero terms. These expressions may then be used to evaluated derivatives of either $W_1$ or $W_2^2$.


\end{document}